\documentclass[fleqn,usenatbib]{mnras} %a4paper

\usepackage{newtxtext,newtxmath}
\usepackage[T1]{fontenc}
\usepackage{ae,aecompl}

\usepackage{graphicx}	% Including figure files
\usepackage{amsmath}	% Advanced maths commands
\usepackage{amssymb}	% Extra maths symbols
\usepackage{grffile}    % allows "dots" in filenames (e.g. m1.5_c3.png)
\usepackage{breqn}
\def\Mdot{\hbox{${\dot M}$}}

\def\km{{\rm\thinspace km}}
\def\s{{\rm\thinspace s}}
\def\yr{{\rm\thinspace yr}}

\def\kmps{\hbox{${\rm\km\s^{-1}\,}$}}

\def\Msol{\hbox{${\rm\thinspace M_{\odot}}$}}
\def\Lsol{\hbox{${\rm\thinspace L_{\odot}}$}}
\def\Msolpyr{\hbox{${\rm\Msol\yr^{-1}\,}$}}
\def\spose#1{\hbox to 0pt{#1\hss}}
\def\ltsimm{\mathrel{\spose{\lower 3pt\hbox{$\sim$}}
        \raise 2.0pt\hbox{$<$}}}
\def\gtsimm{\mathrel{\spose{\lower 3pt\hbox{$\sim$}}
        \raise 2.0pt\hbox{$>$}}}

\title[Non-thermal emission from a colliding wind binary] %max 50 characters
{Colliding-wind binary systems: Diffusive shock acceleration and non-thermal emission}

\author[J.~M.~Pittard et al.]{J.~M.~Pittard$^{1}$\thanks{E-mail:
    j.m.pittard@leeds.ac.uk}, G.~S.~Vila$^{2}$\thanks{Currently at
    CONICET in YPF Tecnolog\'{i}a S.A.} and G.~E.~Romero$^{2}$\\
$^{1}$School of Physics and Astronomy, University of
       Leeds, Woodhouse Lane, Leeds LS2 9JT, UK\\
$^{2}$Instituto Argentino de Radioastronom\'{i}a, CCT-La Plata, CONICET, 1900FWA, La Plata, Argentina\\
}

\date{Accepted ... Received ....; in original form ...}

\pubyear{2020}

\begin{document}
\label{firstpage}
\pagerange{\pageref{firstpage}--\pageref{lastpage}}
\maketitle

% Abstract of the paper
\begin{abstract}
  We present a model for the non-thermal emission from a
  colliding-wind binary. Relativistic protons and electrons are
  assumed to be accelerated through diffusive shock acceleration (DSA)
  at the global shocks bounding the wind-wind collision region.  The
  non-linear effects of the back-reaction due to the cosmic ray
  pressure on the particle acceleration process and the cooling of the
  non-thermal particles as they flow downstream from the shocks are
  included. We explore how the non-thermal particle distribution and
  the keV$-$GeV emission changes with the stellar separation and the
  viewing angle of the system, and with the momentum ratio of the
  winds. We confirm earlier findings that DSA is very efficient when
  magnetic field amplification is not included, leading to
  significantly modified shocks. We also find that the non-thermal
  flux scales with the binary separation in a complicated way and that
  the anisotropic inverse Compton emission shows only a moderate
  variation with viewing angle due to the spatial extent of the
  wind-wind collision.
\end{abstract}

% Select between one and six entries from the list of approved keywords.
% Don't make up new ones.

%%%%%%%%%%%%%%%%%%%%%%%%%%%%%%%%%%%%%%%%%%%%%%%%%%

%%%%%%%%%%%%%%%%% BODY OF PAPER %%%%%%%%%%%%%%%%%%

\section{Introduction}
\label{sec:intro}
Colliding-wind binary (CWB) systems consist of two early-type stars
with powerful winds \citep*[e.g.,][]{Stevens:1992,Pittard:2009}. If
the strength of the winds is not too unbalanced, and/or if the stars
are widely separated, the winds will collide at supersonic speeds
between the stars. This produces a wind-wind collision region (WCR)
where strong global shocks slow the winds and heat the plasma up to
temperatures of $10^{7}$\,K or more.

In some systems the global shocks are collisionless, and are mediated
by magnetic fields rather than through coulombic particle
interactions. This allows particles to undergo diffusive shock
acceleration (DSA), such that a small fraction obtain relativisitic
energies
\citep*[e.g.,][]{Eichler:1993,Benaglia:2003,Dougherty:2003,Reimer:2006,Pittard:2006a,Pittard:2006b}. The
presence of such particles has been confirmed via radio observations
which display a negative spectral index for the flux density
($S_{\nu} \propto \nu^\alpha$, with $\alpha < 0.0$), and which is
  interpreted as synchrotron radiation. In some systems the
non-thermal emission is spatially resolved and is located at the
assumed position of the WCR
\citep*[e.g.,][]{Williams:1997,Dougherty:2000,Dougherty:2005,O'Connor:2005,Dougherty:2006,Ortiz-Leon:2011,Benaglia:2015,Brookes:2016}. In
other systems the non-thermal radio emission is not spatially resolved
but is linked to orbital variability
\citep[e.g.,][]{Blomme:2013,Blomme:2017}.

In contrast to the situation in the radio band, confirmation of non-thermal
X-ray and $\gamma$-ray emission from CWBs has proved extremely
challenging. Until last year the best evidence was a Fermi source
located near to $\eta$~Carinae \citep[e.g.,][]{Reitberger:2015}, an
extreme and unusual CWB composed of an LBV primary in orbit with an as
yet unobserved companion which also has a fast and powerful wind
\citep[e.g.,][]{Pittard:2002,Corcoran:2005,Hamaguchi:2007,Damineli:2008,Okazaki:2008,Parkin:2009,Corcoran:2010,Mehner:2010,Parkin:2011,Madura:2013}.
%Mehner:2010,Parkin:2011,Madura:2013}. 
A second Fermi source is associated with $\gamma^{2}$~Velorum, while
upper limits exist for several other WR+O star CWBs
\citep{Pshirkov:2016}. However, the angular resolution of the Fermi
telescope is relatively poor, and the source circles are large. Thus
it remained possible that the $\gamma$-ray emission detected by Fermi
may actually be coming from other sources than the CWBs \citep{Benaglia:2016}.  

This situation dramatically changed last year when non-thermal X-ray
emission from $\eta$\,Carinae was detected with NuSTAR, a focusing
telescope \citep{Hamaguchi:2018}. These observations narrowed down the
position of the non-thermal emission to within several arc-seconds of
the star, and showed that it varied with the orbital phase of the
binary. In addition, the photon index of the non-thermal X-ray
emission was similar to that found for the $\gamma$-ray spectrum. This
is the conclusive proof that has long been sought, and the NuSTAR
observations provide the crucial support that the detections at X-ray
energies
\citep{Leyder:2008,Sekiguchi:2009,Leyder:2010,Hamaguchi:2014}, GeV
energies
\citep{Tavani:2009,Abdo:2010,Farnier:2011,Reitberger:2012,Reitberger:2015,Balbo:2017}
required. The latest development is the detection of
  $\eta$\,Carinae at energies of 100's GeV by the HESS telescope
  \citep{HESS:2020}.

  In this paper we develop a model for the relativistic particles in
  CWBs and the resulting high-energy non-thermal emission \citep[for
  previous models see,
  e.g.,][]{Dougherty:2003,Pittard:2006a,Pittard:2006b,Reimer:2006,Bednarek:2011,Reitberger:2014a,Reitberger:2014b,Ohm:2015,delPalacio:2016,Reitberger:2017,Grimaldo:2019}. Our
  model is similar to that of \citet{delPalacio:2016} but differs in
  several ways. The most significant difference is that we use the
  semi-analytic model of \citet*{Blasi:2005} to calculate the
  post-shock non-thermal particle distribution. We confirm earlier
  findings that DSA is very efficient when magnetic field
  amplification is not included \citep[e.g.][]{Grimaldo:2019}.  Our
  current focus is the non-thermal X-ray and $\gamma$-ray emission
  that extends up to $10$\,GeV. In Sec.~\ref{sec:model} we describe
  our new model. In Sec.~\ref{sec:results} we present the results and
  we summarize and conclude in Sec.~\ref{sec:summary}.

\section{The model}
\label{sec:model}
To better predict and understand the non-thermal emission from CWBs we
have developed a new, fast and efficient, numerical model. While
models based on hydrodynamical simulations are best able to capture
complex behaviour such as the curvature and skew of the WCR resulting
from orbital dynamics, or the nature of the flow within the WCR, they
are more cumbersome and costly to calculate (especially in
3D). Therefore, there is a place for simpler and faster calculations
that are based on an analytic description of the position of the
contact discontinuity (CD) between the shocked stellar winds. In the
following subsections we describe the geometry of our model, the
acceleration and subsequent cooling of the non-thermal particles in
it, and the non-thermal emission processes that are included in our
calculations. We conclude this section with details about our
``standard model''.

\subsection{The geometry}
\label{sec:geometry}
Our model is based on an axisymmetric description of the WCR in which
it is assumed that the winds collide at constant speeds (we take this
to be the terminal speeds of the winds). Thus, orbital effects and the
acceleration/deceleration of the winds are ignored. Our models are
therefore most appropriate for wide binaries with long orbital periods
where these neglected effects are minimised\footnote{Note that
  \citet{Parkin:2008} developed a simple model for the WCR that {\em
    did} approximate orbital effects.}. We also assume that the global
shocks are coincident with the CD. This is not true in systems where
the cooling length of the shocked plasma is comparable to the stellar
separation (or ``size'' of the WCR), since the shocks stand-off from
the CD in such cases \citep[see, e.g.,][]{Pittard:2018}. However, it
provides a useful first order approximation for the shock positions.

The position of the CD is computed using the equations in
\citet{Canto:1996}. From the apex of the WCR the CD is divided into
segments of 1 degree intervals measured from the secondary star
(hereafter assumed to be the star with the less powerful wind). At
the centre point of each segment the pre-shock wind properties are
calculated: the density, $\rho_{\rm 0}$, and the velocity parallel
($u_{\rm 0\parallel}$) and perpendicular ($u_{\rm 0\perp}$) to the CD.

Each shock segment has two coincident streamlines that flow downstream
along the CD, one for the non-thermal electrons and one for the
non-thermal protons. Each streamline is split into zones. The
size/depth of these zones is controlled by the requirement that the
highest energy particles lose less than 10 per cent of their energy in
any one step (this is why we use two streamlines: the high-energy
non-thermal electrons cool very quickly, which requires small zones,
while the non-thermal protons cool much more slowly and larger zones
can be used). This ensures that the cooling is properly
resolved. There may be many zones per segment. We follow the
post-shock non-thermal particles for a distance of $10\,D$ downstream
of their acceleration point, where $D$ is the stellar separation.

As the particles flow along the streamline they move from the centre
of the current segment towards its edge at a speed of
$u_{\rm 0\parallel}$. If the particles are about to move into the next
segment the timestep is adjusted so that they only just cross into
it. When they cross into the next segment the target photon flux and
post-shock particle density and magnetic field of the new segment
replace the corresponding values from the older segment. In this way
there is a reduction in the rate that the particles cool via inverse
Compton, synchrotron, coulombic and proton-proton cooling, reflecting
the reduction in target photon flux and particle densities along the CD. The
velocity of the flow along the streamline is also updated when the
streamline moves into the next segment, so that the particles
gradually accelerate along their streamline.
 
For the purpose of calculating the emission we gather the particles in
each zone to the centre of the segment that the zone is in. We then
create azimuthal patches by rotating the CD around the
line-of-centres. For the work presented in this paper we create 8
azimuthal patches per CD segment.

\subsection{The diffusive shock acceleration}
\label{sec:DSA}
The main difference to \citet{delPalacio:2016}'s work concerns the
calculation of the non-thermal particle spectrum at each global
shock. \citet{delPalacio:2016} assume that the non-thermal particles
at the two stellar-wind shocks have an energy distribution at
injection of $Q(E) \propto E^{-p}$. The initial post-shock
distribution at each position along each shock is then given by
$N_{0}(E) = Q(E)t_{\rm adv}$, where $t_{\rm adv}$ is the time for the
particles to be advected downstream into the next cell. The
distribution is normalized by the local fraction of the incoming
kinetic energy flux perpendicular to the shock surface that is
converted into non-thermal particles, $f_{\rm NT}$.

In contrast, we solve the diffusion-convection equation for the cosmic
rays using the semi-analytic model of \citet{Blasi:2005} to obtain the
immediate post-shock particle distribution at each shock-segment. The
diffusion of the non-thermal particles is assumed to be energy
dependent in this model (specifically, it is an increasing function of
energy), and is close to B\"{o}hm-like \citep[see Fig.~5
in][]{Blasi:2005}. This means that the spectral index of the particle
distribution, $p$, can also be energy dependent due to the shock
modification process that occurs when DSA is efficient. This is a
major difference to the \citet{delPalacio:2016} model where the
non-thermal particles are assumed to exert no back-reaction on the
thermal plasma.

The \citet{Blasi:2005} shock acceleration model depends on a number of
parameters, such as the pre-shock velocity and Mach number of the flow
normal to the shock ($u_{\rm 0\perp}$ and $M_{\rm 0\perp}$), and the
maximum and injected momenta of the particles ($p_{\rm max}$ and
$p_{\rm inj}$). The latter is set through the parameter
$\chi_{\rm inj}=p_{\rm inj}/p_{\rm th}$ where $p_{\rm th}$ is the
momentum of particles in the thermal peak of the Maxwellian
distribution in the downstream plasma. $M_{\rm 0\perp}$ depends on the
pre-shock gas temperature which we set to $T_{\rm 0}=10^{4}\,$K as
appropriate for photoionized stellar winds. $\chi_{\rm inj}$ is a free
parameter in \citet{Blasi:2005}'s model but, as suggested, we use a
default value of $\chi_{\rm inj}=3.5$.

\citet{Blasi:2005}'s model assumes that the shock is parallel, which
means that the magnetic field is not amplified across the shock. Nor
does the magnetic field become dynamically important. In reality, the
orientation of the pre-shock magnetic field will vary in a complicated
fashion over the WCR. For wide binaries, and where the spin-axis of
each star aligns with the orbital axis, perpendicular shocks are
expected at the apex of the WCR, but other regions of the WCR may have
quasi-parallel shocks. In addition, if the cosmic rays strongly
amplify the magnetic field the far upstream orientation may become
almost irrelevant.  These complications are beyond the current work:
instead, we do not worry about shock obliquity except to determine its
effect on the pre-shock and post-shock velocities normal to the shock,
and we treat the shocks as being parallel along their entire
surfaces. The magnetic field strength in CWBs is also typically very
uncertain, so we treat the pre-shock magnetic field as a
free-parameter in the model: we set its strength through the parameter
$\zeta_{\rm B}=U_{\rm B}/U_{\rm KE}$, where $U_{\rm B}$ and
$U_{\rm KE} = 0.5\rho_{\rm 0}v_{\infty}^2$ are the pre-shock magnetic
and kinetic energy densities, respectively, and we require that
$\zeta_{\rm B} < 1$. The pre-shock magnetic flux density, $B_{\rm 0}$,
is then given by $B_{\rm 0} = \sqrt{8\pi U_{\rm B}}$. Finally, we
assume that the magnetic field is not compressed across the shock,
consistent with our use of \citet{Blasi:2005}'s model and the
assumption that the shock is parallel. In future work we will consider
magnetic field amplification in our model.

Although \citet{Blasi:2005}'s model is for parallel shocks, it does
depend implicitly on the pre-shock magnetic field, which affects the
value of $p_{\rm max}$.  This is because the maximum momentum of the
non-thermal particles, $p_{\rm max}$, is set by the diffusion (escape)
of particles from the shock, where the diffusion length
$l_{\rm diff} = r_{\rm shk}/4$, and where $r_{\rm shk}$ is the
distance of the shock from the star. This gives a maximum proton
energy $E_{\rm max}=l_{\rm diff} eB_{\rm 0} u_{\rm 0\perp}/c$. An
exponential cut-off is then applied to the non-thermal proton spectrum
at $p_{\rm max}$.

The non-thermal electron spectrum has its own maximum
momentum, $p_{\rm max,e}$, which is calculated by balancing the local
acceleration and loss rates, and is similarly truncated at high
energies. Due to the strong inverse Compton cooling in these systems
$p_{\rm max,e} << p_{\rm max}$. The non-thermal electron spectrum is normalized to the non-thermal
proton spectrum by setting $f_{\rm pe} = ep_{\rm ratio}f_{\rm pp}$,
where $f_{\rm pe}$ and $f_{\rm pp}$ are the electron and proton
particle distributions and $ep_{\rm ratio}$ is the ratio of the
electron to proton number density at high energies. The particle
distributions are typically calculated for 140 logarithmic bins in
momentum space from $10^{-6}-10^{8}\,m_{\rm p}c$.

\subsection{The kinetic equation}
The non-thermal particle spectrum downstream of the shock is
calculated by solving the kinetic equation. For a volume co-moving
with the underlying thermal gas, and ignoring diffusion, particle
injection and escape, the energy distribution $n
\equiv dN/dE$ as a function of time $t$ and energy $E$ is given by the
continuity equation \citep*{Ginzburg:1964,Blumenthal:1970}
\begin{equation}
\label{eq:kinetic}
\frac{\partial n(E,t)}{\partial t} + \frac{\partial
  (\dot{E}n(E,t))}{\partial E} = 0.
\end{equation}
The second term is an advection term in energy space due to cooling
processes (e.g. synchrotron, relativistic bremsstrahlung, inverse
Compton - see next subsection). This equation is valid when the energy
losses can be treated as continuous: if the particles lose a large
fraction of their total energy in a single collision then the exact
integro-differential equation must instead be used. However,
significant differences in the electron distributions only occur if
both the electron injection function and the ambient radiation field
are mono-energetic \citep{Zdziarski:1989}.  If the target photon field
is a black-body the continuous energy losss approximation differs by
less than 20 per cent from the exact solution for electron energies $< 10$\,TeV
\citep{Khangulyan:2005}. This difference will be further reduced due
to the increasing importance of the continuous synchrotron energy
losses for the highest energy particles. A continuous treatment for the hadronic
interactions is justified by the large number of inelastic collisions that
allow one to move from the summation to an integral \citep{Stecker:1971}.

If we define the quantity $\tau(E,E')$ as the time required to cool
from an energy $E'$ to $E$ ($\leq E'$),
\begin{equation}
\tau(E,E') = \int_{E}^{E'} \frac{dE''}{\dot{E}(E'')},
\end{equation}
then the evolved distribution at time $t$ of the immediate postshock distribution
$n(E,0)$ is
\begin{equation}
\label{eq:newDistribution}
n(E,t) = \frac{\dot{E}(E')}{\dot{E}(E)} n(E',0),
\end{equation} 
where $E'$ satisfies $\tau(E,E') = t$
\citep[cf.][]{delPalacio:2016}. For a given $E$ and $t$ we determine
$E'$ using a standard numerical root-finding technique. As noted
earlier, Eq.~\ref{eq:newDistribution} is evaluated along the
post-shock streamline at the shorter of the two time intervals for
either the maximum energy of the particles to decrease by 10 per cent
or for the flow to move into the next segment along the CD.

\subsection{Cooling of the downstream non-thermal particles}
\label{sec:cooling}
Post-shock energy losses for the non-thermal electrons occur because
of inverse Compton emission, synchrotron emission, relativistic
bremsstrahlung, coulombic cooling and adiabatic cooling. We assume
that the energy loss is continuous and occurs in a fully ionized
plasma. Further details of the cooling are noted in App.~\ref{sec:AppCooling}.

For the primary electrons, synchrotron cooling is always sub-dominant
to inverse Compton cooling in our models (in the Thomson regime their
relative strength scales as $U_{\rm B}/U_{\rm ph}$). The effect of
adiabatic cooling on the non-thermal electron distribution is usually
seen most strongly at relatively low energies (the high energy
electrons cool rapidly through inverse Compton cooling before they
have had the opportunity to flow very far downstream - see
Figs.~\ref{fig:electron_cooling_timescale}
and~\ref{fig:downstreamElectronMtmDistribution_Model9_WR}).

The non-thermal proton distribution is also subject to cooling as it
flows downstream of the shocks, and in this work we consider
proton-proton pion production and adiabatic expansion, and again
assume that the energy loss is continuous. Since we only consider
acceleration of non-thermal protons in this work, we do not need to
consider the fragmentation of non-thermal nuclei (e.g. He nuclei) due
to collisions with either thermal ions or photons (the latter being
photodisintegration).

\subsection{Non-thermal emission processes}
\label{sec:nt_emission}
For a specified upscattered photon energy we obtain the anisotropic IC
photon flux from each azimuthal patch by integrating over the
non-thermal particle distribution, the incident black-body photon
distribution from each star, and the azimuthal and polar angles of
points on each stellar surface. The anisotropic inverse Compton
emission calculation follows \citet{Cerutti:2007} and
\citet{Vila:2012}\footnote{These theses are available at
  \url{http://citeseerx.ist.psu.edu/viewdoc/download?doi=10.1.1.1019.4046\&rep=rep1\&type=pdf}
  and at \url{https://inis.iaea.org/search/search.aspx?orig_q=RN:46026940}.}, and some details are noted in App.~\ref{sec:appIC}.  Three rotations of
the coordinate system are used to convert a given line of sight into
the coordinate frame used in Fig.~\ref{fig:starElectronObserverGeometry}.

The relativistic bremsstrahlung emission from non-thermal electrons
colliding with thermal protons is calculated using the prescription
noted in App.~\ref{sec:appRB}. The $\gamma$-ray emission from the
decay of neutral pions produced in collisions between thermal and
non-thermal protons is calculated in the delta functional
approximation using the prescription noted in
App.~\ref{sec:appPi0}.

\subsection{Neglected processes}
\label{sec:neglected}
Our focus in this paper is the non-thermal X-ray and $\gamma$-ray
emission up to $10$\,GeV. As a result we do not calculate the
synchrotron emission, which we do not expect to exceed energies
  of $\sim 10^{3}-10^{4}\,$eV (see
  Eq.~\ref{eq:sync_cutoff_energy}). We also do not include
photon-photon absorption (which is inefficient below incident photon
energies of $\sim10-100\,$GeV), or consider emission from the thermal
particles. Finally, we do not consider the formation of, and emission
from, secondary particles.  Each of these processes will be examined
in future work.

\subsection{Standard parameters}
\label{sec:stan_params}
\begin{figure}
\includegraphics[width=8.0cm]{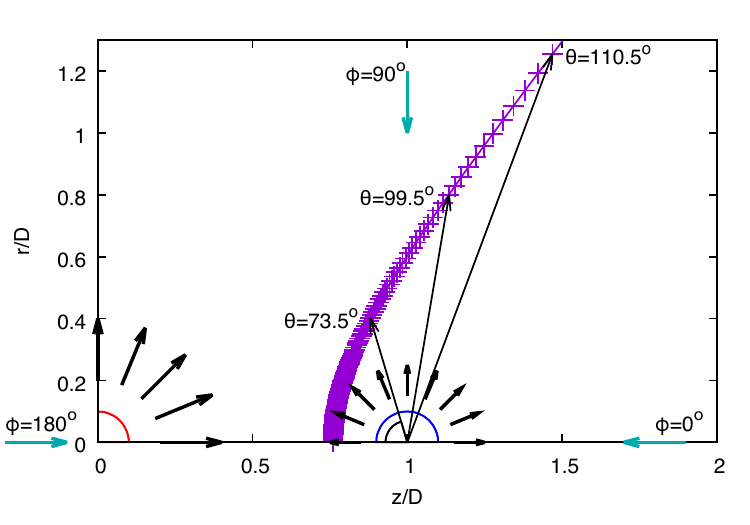}
\caption{The position of the contact discontinuity (CD) in our
  standard model. The primary star is indicated by the red circle, and
  the secondary star by the blue circle. Note that the stars are not
  drawn to scale. The stagnation point of the WCR is at ($z,r$) =
  ($0.74,0.0$)$D$. $\theta$ is the angle between the line of centres
  between the primary and secondary stars and a position on the CD, as
  measured from the secondary star. Marks along the CD indicate the
  centre of segments of $d\theta = 1^{\circ}$ width as seen from the
  secondary star. The mark that is furthest downstream corresponds to
  the 111$^{\rm th}$ segment ($\theta=110.5^{\circ}$ at its
  centre). The viewing angle $\phi$ indicates the angle of the line of
  sight to the observer. The secondary star is in front when
  $\phi=0^{\circ}$, the system is at quadrature when
  $\phi=90^{\circ}$, and the primary star is in front when
  $\phi=180^{\circ}$. For most of our calculations we adopt
  $\phi=90^{\circ}$. For the purpose of our model we assume that the
  global shocks which decelerate each wind are coincident with the CD.
}
\label{fig:CWB_structure}
\end{figure}

In keeping with previous studies
\citep[][]{Dougherty:2003,Pittard:2006a} we examine the emission from
a ``standard'' CWB model of a WR+O system with the parameter values as
noted in Table~\ref{tab:standard_parameters}.
%following parameters: $\Mdot_{\rm WR}=2\times10^{-5} \Msolpyr$,
%$\Mdot_{\rm O}=2\times10^{-6} \Msolpyr$,
%$v_{\rm \infty,WR}=v_{\rm \infty,O}=2000 \kmps$,
%$L_{\rm WR}=2\times10^{5}\,\Lsol$, $L_{\rm O}=5\times10^{5}\,\Lsol$,
%and $D=2\times10^{15}\,$cm. Both stars are assumed to have an
%effective temperature of $40,000$\,K. 
We refer to the WR star as the ``primary'' and the O star as the
``secondary''. The wind momentum ratio,
$\eta = \Mdot_{\rm O}v_{\rm \infty,O}/\Mdot_{\rm WR}v_{\rm \infty,WR} =
0.1$,
and the distance of the stagnation point from the WR and O star is
respectively $r_{\rm WR}=0.74\,D$ and $r_{\rm O} =
0.26\,D$.
Fig.~\ref{fig:CWB_structure} displays the structure of the CD
corresponding to $\eta=0.1$ as in our standard system.

With such parameters the WCR is largely adiabatic, which means that
the global shocks that decelerate each wind stand-off from the CD by
some significant distance. However, we repeat that for the purposes of
this work we assume that the global shocks and the CD are
coincident. We also assume that the winds are composed of pure
hydrogen for the DSA model, but temperatures are calculated assuming
that the average particle mass for both winds is
$\mu = 0.6\,m_{\rm H}$ (i.e. solar abundances). Pre-shock wind
temperatures of $10^{4}$\,K are assumed. The WR star is located at
$(z,r)=(0,0)$ while the O-star is at $(z,r)=(D,0)$. We assume that
$\zeta_{\rm B} = 10^{-3}$ and $\chi_{\rm inj}=3.5$.

The pre-shock density and Mach number of both winds at the stagnation
point are $\rho_{\rm 0} = 2.2\times10^{-19}\,{\rm g\,cm^{-3}}$ and
$M_{\rm 0\perp}=132$. The
pre-shock kinetic energy density, $U_{\rm KE}=0.5\rho_{\rm 0}
v^{2}=4.4\times10^{-3}\,{\rm erg\,cm^{-3}}$. The pre-shock magnetic
energy density, $U_{\rm B} = \zeta_{\rm B}U_{\rm KE} =
4.4\times10^{-6}\,{\rm erg\,cm^{-3}}$, giving $B_{\rm 0}=0.01\,$G.
The total photon energy density (from both stars) at the stagnation point is $U_{\rm ph}=0.021
\,{\rm erg\,cm^{-3}}$. The maximum proton momenta at the WR-star shock
and at the O-star shock are $p_{\rm max} = 8.5\times10^{3}\,{\rm m_{p}c}$ and
$2.7\times10^{3}\,{\rm m_{p}c}$, respectively. The WR-shock
accelerates particles up to higher energies because the incoming wind
has a greater radius of divergence (i.e. it is more planar) than the
O-star wind impinging on the O-star shock. 

The maximum electron Lorentz factor from each shock is
$\gamma_{\rm max,e}\approx 5\times10^{6}$
($p_{\rm max,e}\approx 2700 \,{\rm m_{p}c}$), and is slightly higher for the
WR-star shock than for the O-star shock. In the former case it is
limited by inverse Compton cooling, while in the latter case it is limited
by the maximum energy of the protons. The treatment of $p_{\rm max}$ and
$p_{\rm max,e}$ in the current work is significantly different
compared to our previous work where it was assumed that
$p_{\rm max} = p_{\rm max,e}$, and that these values were the same for
both shocks and along each shock
\citep{Dougherty:2003,Pittard:2006a,Pittard:2006b}. In this sense our new
calculations are more realistic.

For our standard parameters, and with the assumption that the
post-shock magnetic field strength is equal to the pre-shock value, we
find that synchrotron cooling is always sub-dominant to inverse
Compton cooling for the primary electrons - see
Fig.~\ref{fig:electron_cooling_timescale}. However, synchrotron
cooling could become dominant in situations where the magnetic field
is compressed. This will be examined further in future work.

Synchrotron emission also occurs mostly below the energy range that is
of interest to the current work. The synchrotron emission from a
single non-thermal electron cuts off at energies above
\begin{equation}
\label{eq:sync_cutoff_energy}
E = \frac{3h\gamma_{\rm e}^{2}eB\sin\alpha}{4\pi m_{\rm e}c}.
\end{equation}
With $\gamma_{\rm max,e}=5\times10^{6}$ and $\sin\alpha=1$ this gives
$E \approx 4.3\times10^{5}B$\,eV. The pre-shock (and post-shock)
magnetic field strength in our standard model at the apex of the WR-star
and O-star shocks is $B_{\rm 0}=0.01\,$G, which gives $E=4200\,$eV. In
this paper we do not calculate the synchrotron emission (as our focus is
on the keV-GeV energy range).

\begin{table}
\begin{center}
\caption[]{The stellar parameters used in our standard model. Both stars are assumed to have an
effective temperature $T=40,000$\,K. The stellar separation, $D=2\times10^{15}\,$cm.}
\label{tab:standard_parameters}
\begin{tabular}{lll}
\hline
Parameter & WR star & O star\\
\hline
$\Mdot\,\,(\Msolpyr)$ & $2\times10^{-5}$ & $2\times10^{-6}$ \\
$v_{\rm \infty}\,\,(\kmps)$ & $2000$ & $2000$ \\
$L\,\,(\Lsol)$ & $2\times10^{5}$ & $5\times10^{5}$\\
\hline
\end{tabular}
\end{center}
\end{table}

\section{Results}
\label{sec:results} 

We begin by examining various quantities along each shock. We then
examine the distribution of non-thermal particles, and finally
investigate how the predicted emission changes as various parameters
are altered. Unless otherwise noted we adopt our ``standard''
parameters, in which the stellar separation $D = 2\times10^{15}$\,cm,
and the viewing angle $\phi=90^{\circ}$ (i.e. the line-of-sight is
perpendicular to the line-of-centres between the stars - see
Fig.~\ref{fig:CWB_structure}).

\subsection{The standard model}
\subsubsection{Quantities along each shock}
\begin{figure*}
\includegraphics[width=17.5cm]{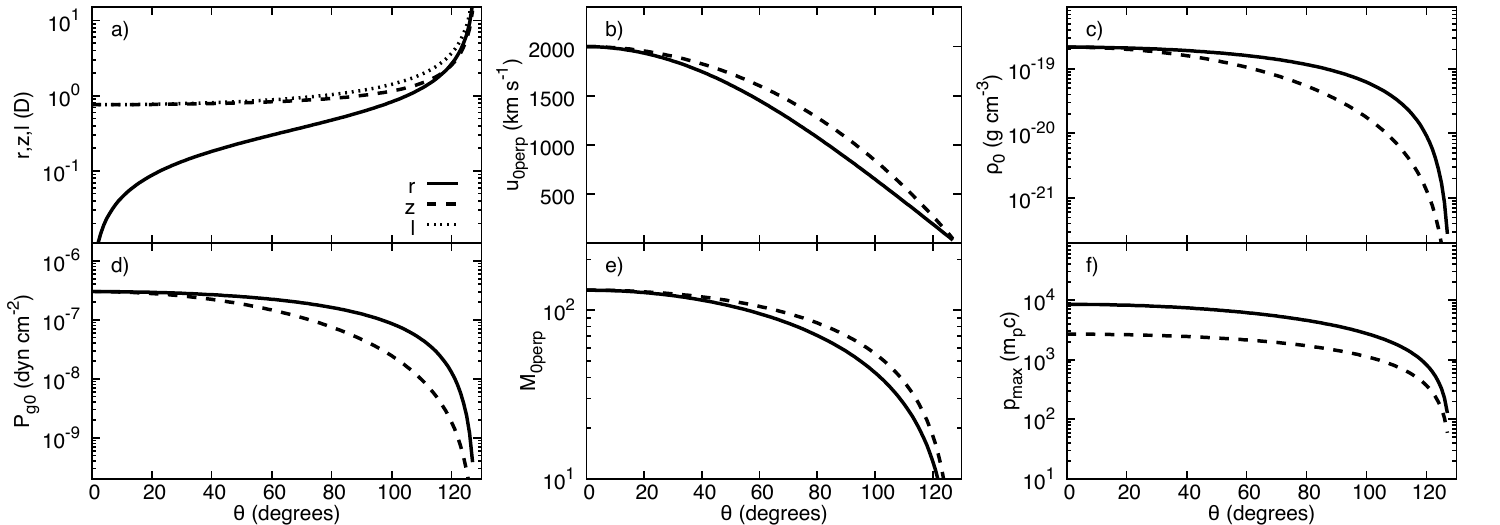}
\caption{Selected quantities along the CD as a function of the angle $\theta$
  from the secondary star. Panel a) shows the $r$ and $z$ position of
the CD segment, and the distance $l$ along the CD. Panels b-e) show
the pre-shock perpendicular wind velocity, density, thermal gas
pressure, and perpendicular Mach number, respectively, while panel f)
shows the maximum non-thermal proton momentum. In panels b-f) the
solid line indicates the properties for the WR-star shock, while the
dashed line indicates the properties for the O-star shock.}
\label{fig:stanCWB_shockstats1}
\end{figure*}

Fig.~\ref{fig:stanCWB_shockstats1} shows various quantities from
our standard model as a function of angle, $\theta$, along the CD as measured
from the secondary star ($\theta=0^{\circ}$ corresponds to the
stagnation point of the WCR on the line-of-centres between the stars,
while $\theta=90^{\circ}$ indicates a point on the CD where
$z=D$). The maximum value of $\theta$ is 180 degrees minus the
half-opening angle of the WCR. For our standard parameters,
$\theta_{\rm max} \approx 130^{\circ}$. $r$, $z$ and $l$, the distance along
the CD from the stagnation point, increase rapidly as $\theta$
approaches its maximum value.

Fig.~\ref{fig:stanCWB_shockstats1}b) shows the perpendicular pre-shock
WR- (solid-line) and O- (dashed line) wind velocity as a function of
$\theta$. At the stagnation point the winds collide head-on and
$u_{\rm 0\perp}$ is equal to the terminal wind speeds. As one moves
off-axis the shocks become gradually more oblique (the WR-shock
becomes more oblique more rapidly), and the perpendicular pre-shock
velocity decreases, reaching zero when $\theta=\theta_{\rm max}$.

Fig.~\ref{fig:stanCWB_shockstats1}c) shows the pre-shock WR-
(solid-line) and O- (dashed line) wind density as a function of
$\theta$. Both densities are identical at the stagnation point
($\rho_{\rm 0} = 2.2\times10^{-19}\,{\rm g\,cm^{-3}}$) due to the fact
that the winds collide at the same speed. The pre-shock WR wind
density falls off more slowly with increasing $\theta$ than the
pre-shock O wind density. Since the pre-shock wind temperatures are
fixed at $10^{4}\,$K, the pre-shock wind pressures in
Fig.~\ref{fig:stanCWB_shockstats1}d) show the same behaviour with
$\theta$ as the pre-shock wind densities. Similarly, the pre-shock
perpendicular Mach number of each wind behaves in the same way as the
pre-shock perpendicular wind speeds (compare
Fig.~\ref{fig:stanCWB_shockstats1}b and~e). The on-axis pre-shock
perpendicular Mach number is $M_{\rm 0\perp} = 132$.

The maximum non-thermal proton momentum at each shock is shown in
Fig.~\ref{fig:stanCWB_shockstats1}f). $p_{\rm max}$ is nearly $10^{4}\,{\rm
  m_{p}c}$ for the WR shock and declines off-axis. The value of $p_{\rm
  max}$ is about 4 times smaller for the O shock due to the reduced
distance of the shock from the star (see Sec.~\ref{sec:DSA}).

\begin{figure*}
\includegraphics[width=17.5cm]{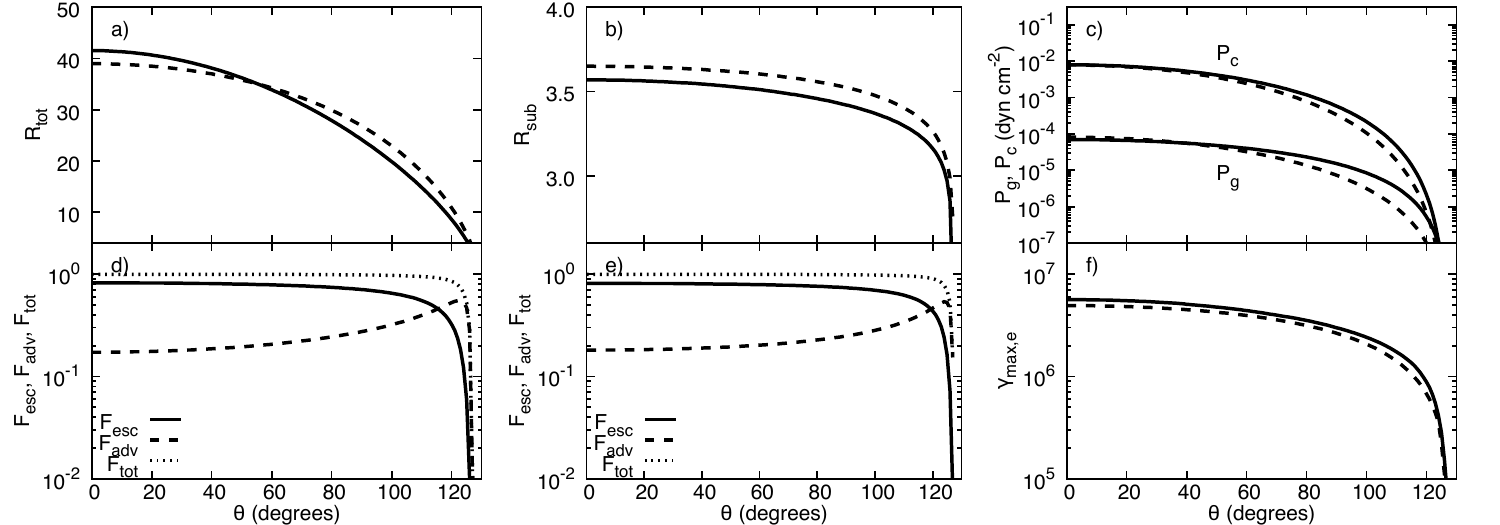}
\caption{Selected quantities along the CD as a function of the angle $\theta$
  from the secondary star. Panels a-c) show the total compression
  ratio of the shock, the compression ratio of the sub-shock, and the
  post-shock pressure from non-thermal ($P_{\rm c}$) and thermal
  ($P_{\rm g}$) particles, respectively. Panel d) shows the advected
  ($F_{\rm adv}$), escaping ($F_{\rm esc}$), and total non-thermal
  particle flux, normalised to the incoming kinetic energy flux, for
  the WR-star shock. Panel e) shows the equivalent for the O-star
  shock. Panel f) shows the maximum Lorentz factor of the non-thermal
  electrons from each shock. In panels a-c) and f) the
solid line indicates the properties for the WR-star shock, while the
dashed line indicates the properties for the O-star shock.}
\label{fig:stanCWB_shockstats2}
\end{figure*}

Fig.~\ref{fig:stanCWB_shockstats1} showed various pre-shock
parameters, including some that are needed for the \citet{Blasi:2005}
DSA model. In Fig.~\ref{fig:stanCWB_shockstats2} we show various
outputs from \citet{Blasi:2005}'s
model. Fig.~\ref{fig:stanCWB_shockstats2}a) shows $R_{\rm tot}$, the
shock total compression ratio. Strong shocks in gas with a ratio of
specific heats $\gamma=5/3$ have a compression ratio of 4, but
$R_{\rm tot}$ can increase significantly when DSA efficiently
accelerates non-thermal particles that then escape upstream from the
shock. This is indeed the case in our standard model, where we see
that $R_{\rm tot}$ reaches values of order 40\footnote{Less extreme
  compression ratios are achieved when the dynamical feedback of the
  magnetic field amplification is considered \citep[see,
  e.g.,][]{Caprioli:2009}.}. The lower value of $p_{\rm max}$ on the
line-of-centres for the O shock causes $R_{\rm tot}$ to be slightly
lower than for the WR shock.  $R_{\rm tot}$ decreases with increasing
$\theta$ as the shocks become more oblique, and $u_{\rm 0\perp}$,
$M_{\rm 0\perp}$ and $p_{\rm max}$ all decline.

Fig.~\ref{fig:stanCWB_shockstats2}b) shows the compression ratio
across the sub-shock, $R_{\rm sub}$. The sub-shock is a discontinuity in the
overall shock structure. $R_{\rm sub}$ is $\approx 3.5$ for both the WR and O
shock and decreases slightly as $\theta$ increases, before falling
rapidly as $\theta \rightarrow \theta_{\rm max}$. The sub-shock, plus
any shock-precursor, is responsible for heating the thermal
plasma.

The post-shock thermal, $P_{\rm g}$, and non-thermal, $P_{\rm c}$,
particle pressure is shown in Fig.~\ref{fig:stanCWB_shockstats2}c). It
is clear that $P_{\rm c}$ exceeds $P_{\rm g}$ by a factor of 100 at
the shock apex. This difference reduces as $\theta$ increases, until
at large $\theta$ the value of $P_{\rm c}$ drops to a value similar to
that of $P_{\rm g}$ as DSA becomes less efficient.

In Fig.~\ref{fig:stanCWB_shockstats2}d) we see the variation with
$\theta$ of the fraction of the incoming WR-wind kinetic energy flux that is
advected downstream in non-thermal particles, $F_{\rm adv}$. Also shown is the
fraction that is carried upstream by escaping non-thermal particles,
$F_{\rm esc}$, and the total non-thermal particle flux ($F_{\rm tot} =
F_{\rm adv} + F_{\rm esc}$). Due to the efficient DSA that occurs over
most of the shocks, $F_{\rm tot}\approx 1.0$, and $F_{\rm esc} >
F_{\rm adv}$. Only once $\theta \gtsimm 90^{\circ}$ does the
efficiency drop. At $\theta=0^{\circ}$, $F_{\rm tot}=0.992$, while
$F_{\rm tot} = 0.9$ and $0.5$ at $\theta=116^{\circ}$ and
$125^{\circ}$, respectively. Fig. ~\ref{fig:stanCWB_shockstats2}e)
shows the same quantities for the O shock. The same general behaviour
is seen, though the shock stays efficient out to slightly higher
values of $\theta$ (in this case $F_{\rm tot} = 0.9$ and $0.5$ at $\theta=120^{\circ}$ and
$126^{\circ}$). Our results can be compared against Fig.~10 in
\citet{Blasi:2005} where these quantities are shown as a function of
the shock Mach number.

The maximum electron Lorentz factor is shown in
Fig.~\ref{fig:stanCWB_shockstats2}f) for the two shocks. Both shocks
have values of $\gamma_{\rm max,e}\approx 5\times10^{6}$ on the
line-of-centres, and this value drops only slightly as $\theta$
increases. Only once past $\theta\approx100^{\circ}$ does it begin to
drop more rapidly. Thus the assumption of a constant value of
$\gamma_{\rm max,e}$ in our previous work
\citep{Dougherty:2003,Pittard:2006a,Pittard:2006b} was a reasonably
good one. The value of $\gamma_{\rm max,e}$ in our models is strongly
dependent on $D$ and $B_{\rm 0}$.

\subsubsection{The particle distributions}
\begin{figure}
\includegraphics[width=8.0cm]{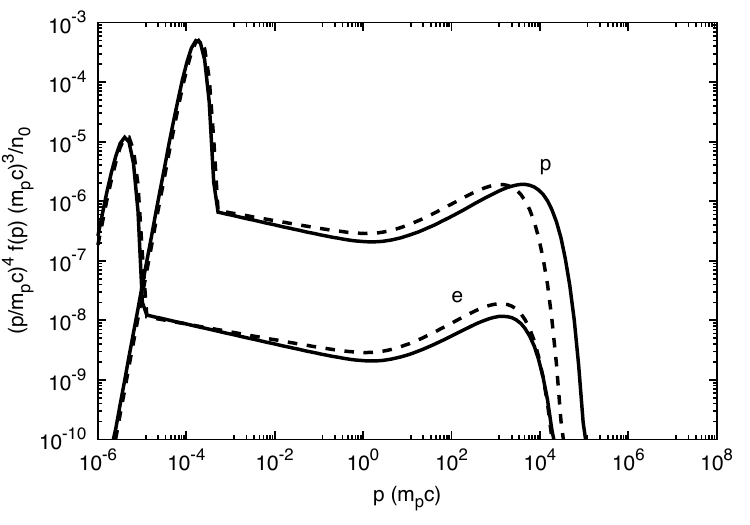}
\caption{The proton and electron distributions for the WR
  shock (solid line) and O shock (dashed line) for
  $\theta=0^{\circ}$. For both shocks $n_{0} = 1.3\times10^{5}\,{\rm
    cm^{-3}}$. The thermal peaks are visible at low momenta.}
\label{fig:stanCWB_mtmSpectra}
\end{figure}

\begin{figure}
\includegraphics[width=8.0cm]{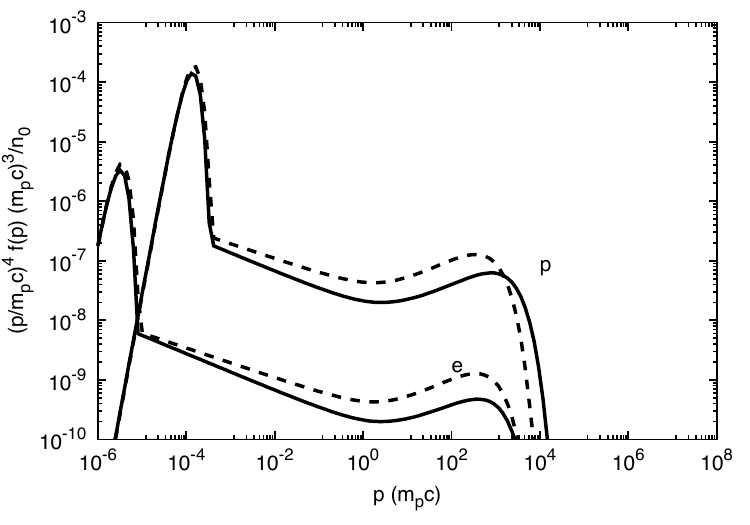}
\caption{The proton and electron distributions for the WR
  shock (solid line) and O shock (dashed line) for
  $\theta=110^{\circ}$. For the WR shock $n_{0} =
  2.0\times10^{4}\,{\rm cm^{-3}}$, while for the O shock $n_{0} = 4.2\times10^{3}\,{\rm cm^{-3}}$.}
\label{fig:stanCWB_mtmSpectra2}
\end{figure}

Figs.~\ref{fig:stanCWB_mtmSpectra} and~\ref{fig:stanCWB_mtmSpectra2}
show the distributions of the thermal and non-thermal particles immediately
downstream of the shock. In each figure the top two lines are the
proton distributions, while the bottom two are the electron
distributions. The particle distributions are shown for the WR shock
(solid line) and the O shock (dashed
line). Fig.~\ref{fig:stanCWB_mtmSpectra} shows the distributions for $\theta=0^{\circ}$,
while Fig.~\ref{fig:stanCWB_mtmSpectra2} shows them for
$\theta=110^{\circ}$. In all cases the distributions clearly show
strong shock modification, with most of the energy pushed towards the
highest momenta. As previously noted, the maximum proton momentum is lower at the O shock
than at the WR shock. The positions of the thermal peak also indicates
the effect of cooler downstream thermal particles for modified shocks.

\begin{figure}
\includegraphics[width=8.0cm]{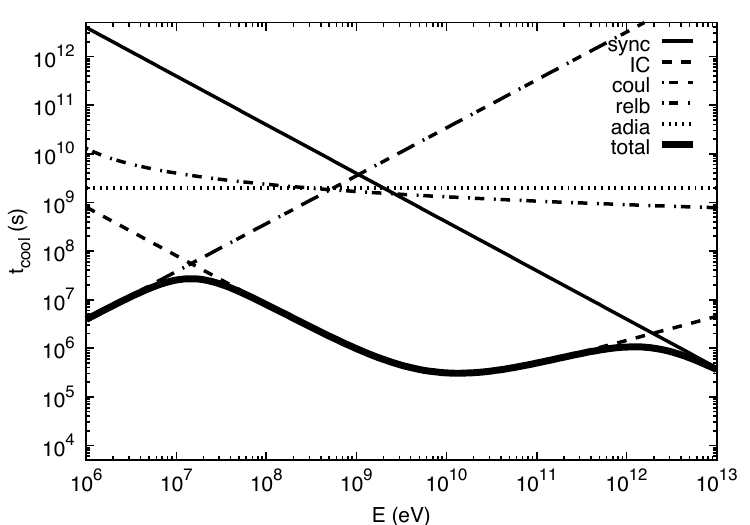}
\caption{The cooling time $t_{\rm cool} = E/|\dot{E}|$ as a function of
  electron energy for electrons located immediately post-WR-shock for
  $\theta=0.5^{\circ}$ in the ``standard'' model with $D =
  2\times10^{15}\,$cm. The parameters are $n_{\rm
    e}=5\times10^{5}\,{\rm cm^{-3}}$, $T = 4\times10^{4}$\,K,
  $B = 0.01$\,G, $R = 1.5\times10^{15}$\,cm, $v = v_{\rm CD} =
  0.8\,\kmps$. Coulomb cooling dominates at low energies, synchrotron
  cooling dominates at high energies, and inverse Compton cooling
  dominates at intermediate (10\,MeV$ - $2\,TeV) energies.
}
\label{fig:electron_cooling_timescale}
\end{figure}

The cooling timescale for electrons ($t_{\rm cool} = E/|\dot{E}|$) for
the various energy-loss mechanisms is shown in
Fig.~\ref{fig:electron_cooling_timescale}. The electrons are assumed
to be located near the apex of the WCR (specifically, they are
immediately behind the WR shock at $\theta=0.5^{\circ}$). The cooling
time is dominated by different mechanisms in different energy
ranges. At the lowest energies, losses due to Coulomb scattering
dominate, whereas inverse Compton cooling takes over for
$E \gtsimm 10$\,MeV. At the highest energies shown inverse Compton
cooling loses its dominance as its cross section reduces and
synchrotron cooling becomes dominant at energies above $\sim
1$\,TeV. The relative strength of the inverse Compton and synchrotron cooling
depends on a number of the model parameters, including $L_{\rm WR}$,
$L_{\rm O}$, $D$, the pre-shock magnetic field, $B_{\rm 0}$, and the
amount of compression/amplification of the magnetic field.

\begin{figure}
\includegraphics[width=8.0cm]{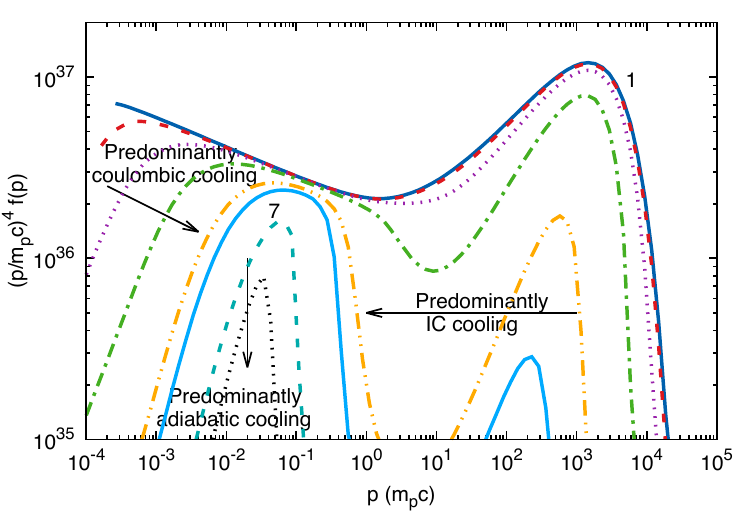}
\caption{The downstream cooling of the electron distribution of the WR
  shock for $\theta=0.5^{\circ}$. The immediate post-shock distribution is the top-most line
  (labelled ``1''), and the distribution shifts downwards and inwards with increasing
  cooling. Some properties of each distributions are noted in Table~\ref{tab:downstreamElectronMtmDistribution_Model9_WR}. $D = 2\times10^{15}\,$cm. }
\label{fig:downstreamElectronMtmDistribution_Model9_WR}
\end{figure}

The downstream cooling of the non-thermal electron distribution from
the WR shock at $\theta=0.5^{\circ}$ is shown in
Fig.~\ref{fig:downstreamElectronMtmDistribution_Model9_WR}. Inverse
Compton cooling dominates the cooling of the high energy electrons,
while coulombic cooling dominates at low energies. The properties of
the distributions are noted in
Table~\ref{tab:downstreamElectronMtmDistribution_Model9_WR}. Because
the post-shock tangential velocity is low (the wind collides almost
normal to the shock) it takes a long time for the streamline to
increase its value of $\theta$ (which it can do only in $1^{\circ}$
steps). By the time of the final distribution shown the particles have
flowed downstream a total distance of $0.015\,D$, taking
$1.3\times10^{7}$\,s to do so. In the code, the final distribution
shown is actually the 158$^{\rm th}$ distribution stored along this
streamline (i.e., the cooling is resolved very well), and a total of
172 distributions are calculated and stored along this
streamline. 

\begin{table}
\begin{center}
  \caption[]{The properties of the distributions shown in
    Fig.~\ref{fig:downstreamElectronMtmDistribution_Model9_WR}. The
    distributions are numbered from 1 to 8, with the amount of cooling
    increasing with the distribution index. The value of $\theta$, the
    arc-length along the CD from the stagnation point, and the elapsed
    time since the shock are noted.}
\label{tab:downstreamElectronMtmDistribution_Model9_WR}
\begin{tabular}{llll}
\hline
Index & $\theta$ ($^{\circ}$) & $l/D$ & $t$\,(s) \\
\hline
1 & 0.5 & $1.0\times10^{-6}$ & 2670 \\
2 & 0.5 & $6.5\times10^{-6}$ & $1.7\times10^{4}$ \\
3 & 0.5 & $3.6\times10^{-5}$ & $9.4\times10^{4}$ \\
4 & 0.5 & $1.5\times10^{-4}$ & $4.0\times10^{5}$ \\
5 & 0.5 & $6.1\times10^{-4}$ & $1.6\times10^{6}$ \\
6 & 0.5 & $9.6\times10^{-4}$ & $2.5\times10^{6}$ \\
7 & 1.5 & $4.3\times10^{-3}$ & $7.4\times10^{6}$ \\
8 & 3.5 & 0.015 & $1.3\times10^{7}$ \\
\hline
\end{tabular}
\end{center}
\end{table}

\subsubsection{The non-thermal emission}
\begin{figure}
\includegraphics[width=8.0cm]{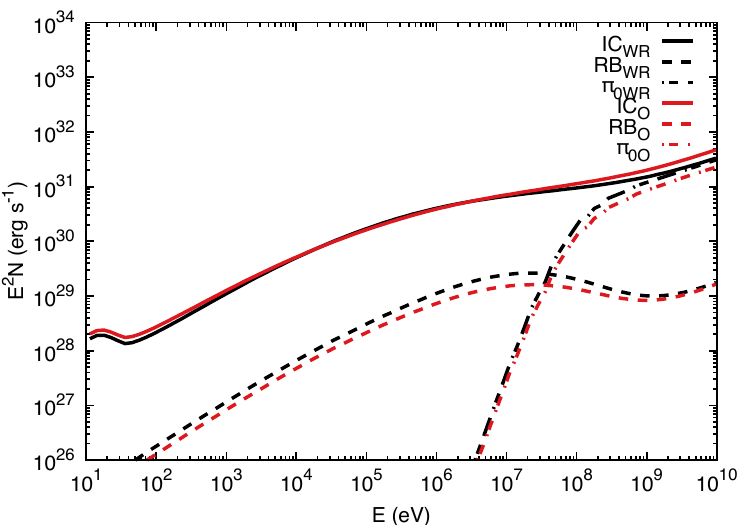}
\caption{The non-thermal emission from each shock from our standard
  model. $D=2\times10^{15}\,{\rm cm}$ and $\phi=90^{\circ}$. The
    black lines show the emission from the WR-star shock, while the red lines show
    it from the O-star shock.}
\label{fig:whichShockDominates_dsep2e15}
\end{figure}

Fig.~\ref{fig:whichShockDominates_dsep2e15} shows the non-thermal
emission from our standard model. The inverse Compton emission is
dominant for $E \ltsimm 1$\,GeV, while $\pi^{0}$-decay emission
becomes comparable in strength at higher energies. The emission from
relativistic bremsstrahlung is always at least an order of magnitude
fainter than the inverse Compton emission. Both shocks contribute
roughly equally to the inverse Compton emission, though the
relativistic bremsstrahlung and $\pi^{0}$-decay emission from the
WR-star shock is about twice as bright as that from the O-star
shock. The O-star shock has brighter inverse Compton emission at
energies above 10\,MeV than the WR-star shock. The signature of strong
shock modification is visible as the curvature in the inverse Compton
and relativistic bremsstrahlung emission which leads to increased flux
at higher energies. Detection of such curvature from actual systems
would indicate strong shock modification.

\subsection{Effect of binary separation and downstream cooling}
We now examine how the non-thermal particle distributions and the
resulting emission changes when the stellar separation, $D$, is altered.

\subsubsection{Expected scaling}
If the cooling timescale of the post-shock thermal particles
($t_{\rm cool,th}$) is long compared to their dynamical timescale to
flow out of the system ($t_{\rm dyn}$), then the WCR behaves
self-similarly, and its volume $V$ scales as $D^{3}$. In such
circumstances the total emission from thermal particles, with number
density $n_{\rm th}$, scales as $L_{\rm th} \propto n_{\rm th}^{2} V \propto D^{-1}$
\citep{Stevens:1992}.

We now consider how the non-thermal emission should scale. The
non-thermal particle density, $n_{\rm NT}$, scales as $D^{-2}$. If the
non-thermal particles also do not strongly cool (i.e. their cooling
timescale $t_{\rm cool,NT} \ltsimm t_{\rm dyn}$), then they fill the
WCR, and so the volume that they occupy scales as $D^{3}$. For the
inverse Compton emission, the number density of stellar photons,
$n_{\rm ph}$, also scales as $D^{-2}$, so we expect
$L_{\rm IC} \propto n_{\rm NT} n_{\rm ph} V \propto D^{-1}$. We also
expect the relativistic bremsstrahlung and the $\pi^{0}$-decay
emission to both scale as $n_{\rm th} n_{\rm NT} V \propto D^{-1}$.
%, so that $L_{\rm RB}\propto D^{-1}$ and $L_{\rm \pi^{0}}\propto D^{-1}$.

Now consider the situation where the thermal gas in the WCR is largely
adiabatic but where there is very rapid cooling of the non-thermal
particles. As noted by \citet{Hamaguchi:2018}, the cooling length
$l_{\rm cool,NT} \propto D^{2}$, so the ``volume'' that the
non-thermal electrons occupy prior to being cooled below some energy
limit ($V_{\rm NT} < V$) is given by the area of the shocks
($A \propto D^{2}$) multiplied by the cooling length: i.e.
$V_{\rm NT} = A\,l_{\rm cool,NT} \propto D^{4}$. In such cases we
expect the non-thermal emission to scale as $D^{0}$.

We expect $p_{\rm max}$ to be independent of $D$, since
$p_{\rm max} \propto r_{\rm shk} B_{\rm 0}$, with
$r_{\rm shk} \propto D$ and
$B_{\rm 0} \propto U_{\rm B}^{1/2} \propto U_{\rm KE}^{1/2} \propto
\rho_{\rm 0}^{1/2} \propto D^{-1}$.
On the other hand, $p_{\rm max,e}$, may depend on the strength of the
inverse Compton cooling. By balancing the rate of energy gain through
DSA with the rate of energy loss through inverse Compton cooling, we
find that in such cases
$p_{\rm max,e} \propto r_{\rm shk} B_{0}^{1/2} \propto D^{1/2}$
\citep[see, e.g.,][]{Pittard:2006a}. We find the same scaling of
$p_{\rm max,e}$ with $D$ if instead $p_{\rm max,e}$ depends on
synchrotron cooling.

\subsubsection{The particle distributions}
\begin{figure}
\includegraphics[width=8.0cm]{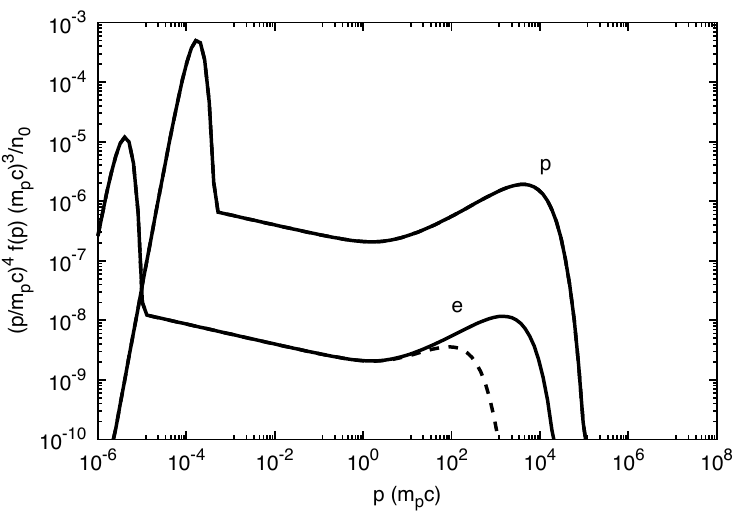}
\caption{The proton and electron distributions for the on-axis WR
  shock as a function of $D$. The solid lines have $D =
  2\times10^{15}\,$cm, while the dashed lines have $D =
  2\times10^{14}\,$cm. The solid and dashed lines are coincident for the proton
  distributions when normalised by the pre-shock number density.}
\label{fig:stanCWB_mtmSpectra_varDsep}
\end{figure}

Fig.~\ref{fig:stanCWB_mtmSpectra_varDsep} compares the on-axis
($\theta=0^{\circ}$) post-shock particle distributions for the WR
shock for $D = 2\times10^{14}\,$cm and $D = 2\times10^{15}\,$cm,
normalized to the pre-shock number density. Because $M_{0\perp}$,
$u_{0\perp}$ and $p_{\rm max}$ are all independent of $D$, the
(normalized) proton distributions are identical for the two distances.
However, the electron distribution is cut off at a lower maximum
momentum when $D = 2\times10^{14}\,$cm due to the enhanced inverse
Compton cooling. We find that $p_{\rm max} \propto D^{0}$ and $p_{\rm
  max,e} \propto D^{1/2}$ as expected.

\subsubsection{The non-thermal emission}
\begin{figure}
\includegraphics[width=8.0cm]{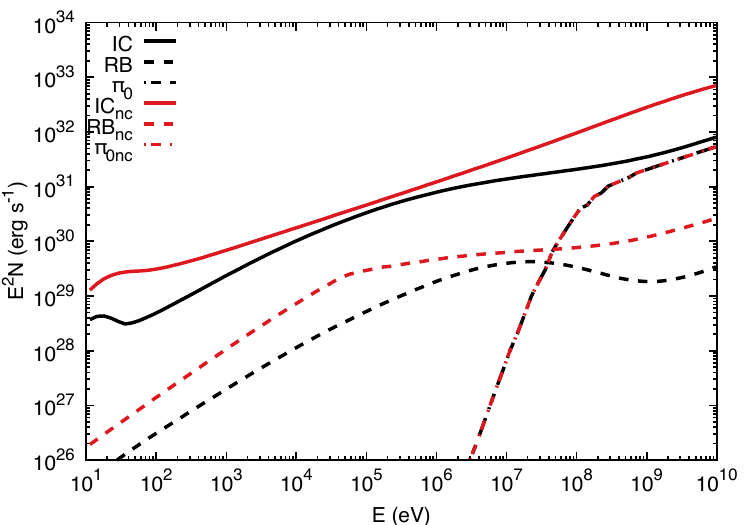}
\caption{The effect of cooling on the downstream non-thermal
  particles and their subsequent emission. The red lines include only
  adiabatic cooling, while the black lines include all cooling
  processes considered in this work (adiabatic, coulombic, inverse
  Compton, synchrotron and relativistic bremsstrahlung cooling for the electrons;
  adiabatic and proton-proton cooling for the protons). $D=2\times10^{15}\,{\rm cm}$
  and $\phi=90^{\circ}$.}
\label{fig:effectOfCooling_dsep2e15}
\end{figure}

Before we examine the effect on the non-thermal emission of varying
$D$, it is helpful to examine the effect of downstream cooling on the
non-thermal emission for our standard model
($D=2\times10^{15}\,{\rm cm}$). This is indicated in
Fig.~\ref{fig:effectOfCooling_dsep2e15}, where the difference between
the red and black lines shows the effect of including all the cooling
processes noted in Sec.~\ref{sec:cooling} versus considering only
adiabatic cooling).  The high energy electrons cool strongly due to IC
emission while the lower energy electrons cool through coulombic
collisions. Cooling of the non-thermal electrons reduces the inverse
Compton and relativistic bremsstrahlung emission at GeV energies by
$\sim 1$\,dex. In contrast, there is little cooling of the non-thermal
protons, as evidenced by the almost unchanged $\pi^{0}$-decay
emission.

The inverse Compton emission from particles along the single
$\theta=0.5^{\circ}$ streamline from the WR-shock is approximately
flat in the $E^{2} N$ spectrum at 10\,MeV, with an upward curvature
with increasing energy due to the strongly modified nature of the
shock. In contrast, Fig.~\ref{fig:effectOfCooling_dsep2e15} shows that
there is a slight rise in the inverse Compton emission at 10\,MeV (see
the solid black line).  This rise is caused by higher energy breaks in
the electron distributions along the other streamlines.
 
\begin{figure}
\includegraphics[width=8.0cm]{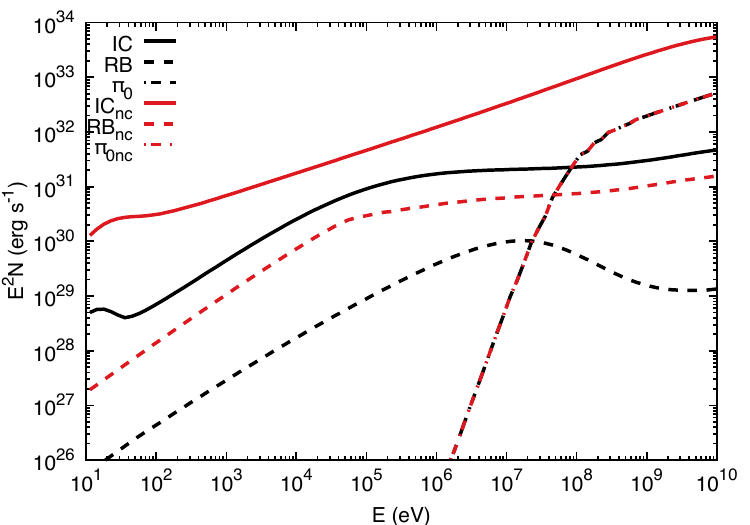}
\caption{As Fig.~\ref{fig:effectOfCooling_dsep2e15} but for
  $D=2\times10^{14}\,{\rm cm}$. $\phi=90^{\circ}$.}
\label{fig:effectOfCooling_dsep2e14}
\end{figure}

The effect of downstream cooling on the non-thermal emission for a
model with reduced binary separation ($D=2\times10^{14}\,{\rm cm}$) is
shown in Fig.~\ref{fig:effectOfCooling_dsep2e14}. Compared to
Fig.~\ref{fig:effectOfCooling_dsep2e15} we see that the effect of
cooling has strengthened, as expected given the reduced separation.

\begin{figure}
\includegraphics[width=8.0cm]{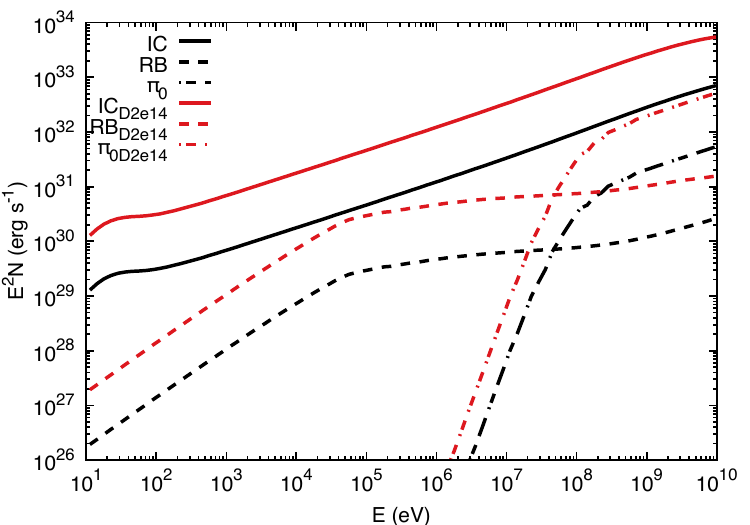}
\caption{The effect of binary separation on the non-thermal
  emission. Only adiabatic cooling of the downstream non-thermal
  particles has been applied. The black lines are for
  $D=2\times10^{15}$\,cm and the red lines are for $D=2\times10^{14}$\,cm. $\phi=90^{\circ}$.}
\label{fig:comparisonOfDistance_nocooling}
\end{figure}

Fig.~\ref{fig:comparisonOfDistance_nocooling} shows the effect of
binary separation on the non-thermal emission if only adiabatic cooling is applied
to the downsteam non-thermal particles. We see that all emission
processes scale as $D^{-1}$, as expected.

\begin{figure}
\includegraphics[width=8.0cm]{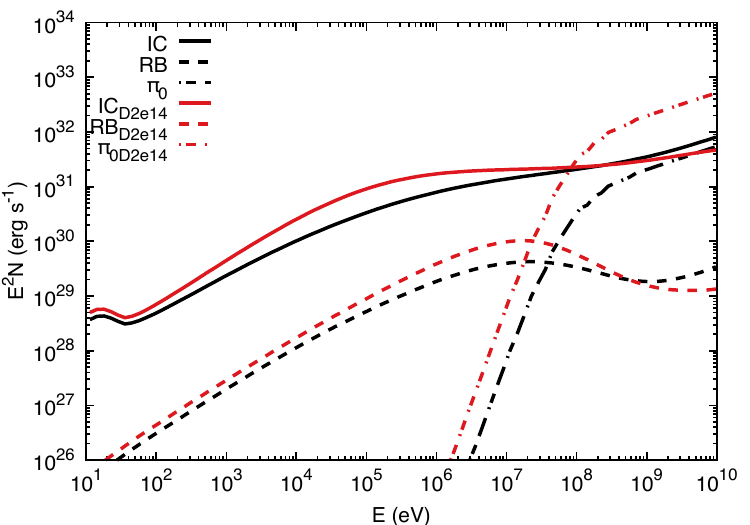}
\caption{The effect of binary separation on the non-thermal emission. 
Cooling of the downstream non-thermal particles has been fully applied. The black lines are for
  $D=2\times10^{15}$\,cm and the red lines are for $D=2\times10^{14}$\,cm. $\phi=90^{\circ}$.}
\label{fig:comparisonOfDistance_withcooling}
\end{figure}

Finally, Fig.~\ref{fig:comparisonOfDistance_withcooling} shows the
effect of binary separation on the non-thermal emission if cooling is
fully applied to the downsteam non-thermal particles. We now find that the
previous $D^{-1}$ scaling of the inverse Compton and relativistic
bremsstrahlung emission disappears and the change with $D$ becomes
much reduced. However, the emission from $\pi^{0}$-decay still varies strongly (and
almost as $D^{-1}$), again illustrating that the non-thermal protons
do not undergo strong downstream cooling.

This behaviour contrasts with some other modelling work in the
literature. For instance, Figs.~12 and~13 in \citet{Reimer:2006} show
the relativistic bremsstrahlung and $\pi^{0}$-decay $\gamma$-ray
spectra {\it from the acceleration region} in their model scaling
roughly as $D^{-4}$. Unfortunately the scaling behaviour of the total
emission (from the acceleration and convection region combined) is not
shown. Likewise, the flux variations in Models A-C in Fig.~10 of
\citet{Reitberger:2014b} show no sign of becoming independent of $D$,
despite the close-ish separations. We note that Reitberger et al. use
a fixed size for their computational volume. A plausible explanation
for their results is that a a greater fraction of the total emission
was ``missed'' from the model with the wider stellar separation.

\begin{figure}
\includegraphics[width=8.0cm]{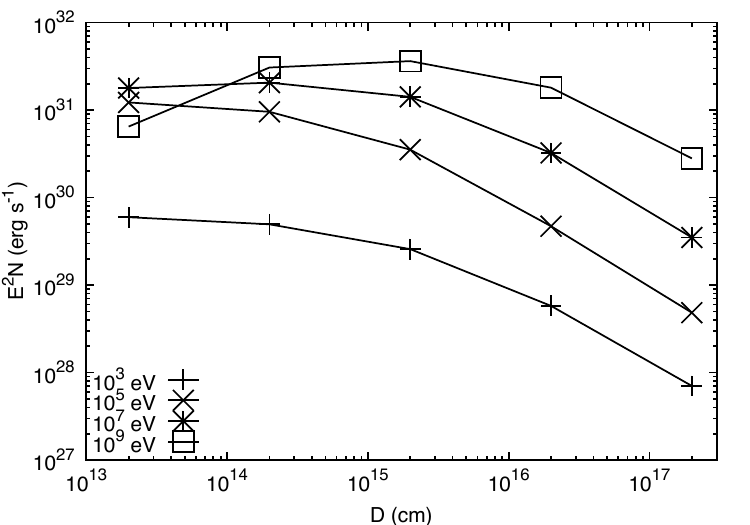}
\caption{The effect of binary separation on the inverse Compton
  emission. Data points are shown at four energies: $10^{3}$,
  $10^{5}$, $10^{7}$, and $10^{9}$\,eV. $\phi=90^{\circ}$.}
\label{fig:comparisonOfDistance_withcooling_IC_v2}
\end{figure}

Fig.~\ref{fig:comparisonOfDistance_withcooling_IC_v2} shows how the
inverse Compton emission changes with stellar separation. At low
energies and large separations the slope of the lines is $\approx -1$,
indicating that the responsible particles fill the WCR (i.e. they are
not strongly cooling as they flow away from the shock). However,
cooling becomes increasingly important as $D$ decreases. The emission
at $10^{3}$ and $10^{5}$\,eV no longer scales as $D^{-1}$, but still
scales to some inverse power. In contrast, at the higher energies we
find that the flux reaches a maximum at an intermediate value of $D$,
and then decreases as $D$ becomes still smaller. This is caused by
$p_{\rm max,e}$ decreasing with decreasing $D$, which has the knock-on
effect that the fluxes begin to be affected by the exponential cut-off
of the non-thermal electron particle distribution.

\begin{figure}
\includegraphics[width=8.0cm]{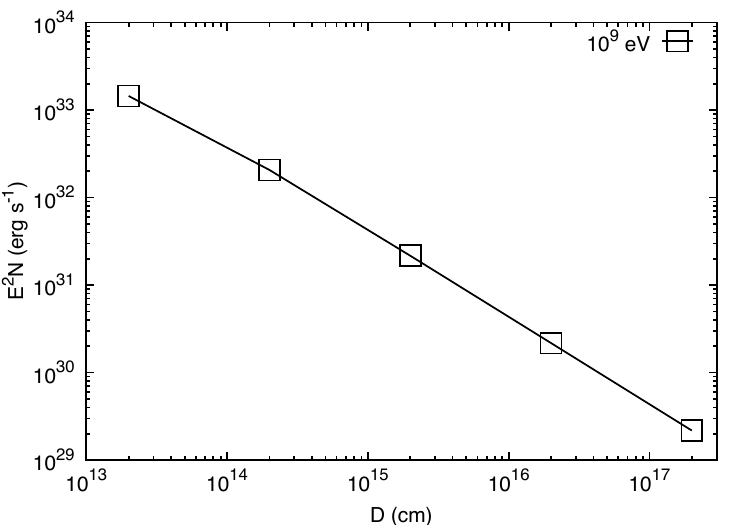}
\caption{The effect of the binary separation on the $\pi^{0}$-decay
  emission at $10^{9}$\,eV.}
\label{fig:comparisonOfDistance_withcooling_Pi0_v2}
\end{figure}

Fig.~\ref{fig:comparisonOfDistance_withcooling_Pi0_v2} shows the
effect of changing the binary separation on the $\pi^{0}$-decay emission at
$10^{9}$\,eV. At large $D$ we again see that the flux scales as
$D^{-1}$, but again witness a turndown in this slope as $D$
decreases. It is clear that cooling of the downstream non-thermal
protons starts to become significant for $D \ltsimm 10^{14}$\,cm.

\subsubsection{Comparison to observations}
At the time of writing the strongest evidence for orbital variability
of non-thermal X-ray and $\gamma$-ray emission comes from Fermi
observations of $\eta$\,Carinae. \citet{Balbo:2017} find that of the
two emission components seen by Fermi, the flux of the low-energy ($0.3-10$\,GeV) 
component is modulated by the orbit, being stronger
near periastron and weaker at apastron. Overall, it varies by less
than a factor of 2. This component is likely inverse Compton emission,
and it is probably not significantly affected by photon-photon
absorption. On the other hand, the high-energy ($10-300$\,GeV) component
varies by a factor of $3-4$ and is different during
the two periastrons that are observed (see their Fig.~5). This
component is likely emission from $\pi^{0}$-decay and will be strongly
affected by photon-photon absorption. 

In contrast, Fig.~\ref{fig:comparisonOfDistance_withcooling_IC_v2}
shows that at $E=10^{9}$\,eV, the inverse Compton flux in our model
{\em increases} by a factor of $\approx 5$ when $D$ increases from
$2-20\times10^{13}\,$cm (for $\eta$\,Car, $D=2.3-44\times10^{13}\,$cm
between periastron and apastron). Thus the flux is our model behaves
the opposite way to the observed emission from $\eta$\,Car, which
{\em decreases} with increasing $D$. It will be interesting to see if
these differences can be reconciled with a dedicated application of
our model to $\eta$\,Car (the modelling in \citet{Balbo:2017} is able
to reproduce the variation, to first order).

\subsection{Effect of observing angle}
\begin{figure}
\includegraphics[width=8.0cm]{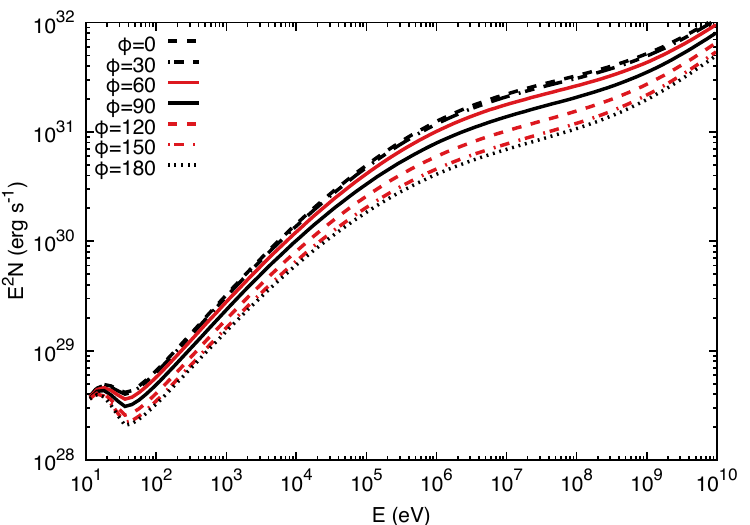}
\caption{The effect of the observing angle on the inverse Compton
  emission. The standard model has $\phi=90^{\circ}$. At
  $\phi=0^\circ$ the secondary star is in front, while the primary
  star is in front when $\phi=180^\circ$. $D=2\times10^{15}$\,cm.}
\label{fig:effectOfObservingAngle}
\end{figure}

We now examine the effect on the non-thermal emission of changing the
observing angle. Because no absorption processes are included in the
current model only the anisotropic nature of the inverse Compton
emission affects the observed non-thermal emission. This is shown in
Fig.~\ref{fig:effectOfObservingAngle}. Since our model is
axisymmetric, changing only the observing angle covers any orientation
of the system relative to the observer. At an observing angle
$\phi=0^\circ$ the secondary star is in front, quadrature is at
$\phi=90^\circ$, and the primary star is in front at
$\phi=180^\circ$. The strongest emission occurs when the secondary
star is in front, while the weakest emission occurs when the primary
star is in front. This agrees with expectations since the secondary
star is the major source of incident photons and is closest to the
WCR.

There is not much change in the emission when one of the stars is
within $\sim30^\circ$ from being directly in front. This is likely due
to the fact that the asymptotic half-opening angle of the WCR for
$\eta=0.1$ is $\approx 50^\circ$ \citep[][]{Pittard:2018}, so that the
line of sight is still within the shock cone for this range of viewing
angle. We find that it is only when the line of sight moves outside of
the shock cone that the emission become more sensitive to changes in
viewing angle.

The overall variation is about a factor of 3.5 (as measured at
$E=1 \,$MeV). This is much lower than in other work
\citep[e.g.][]{Reimer:2006,Dubus:2008} which we attribute to the way
that the WCR wraps around the secondary star in our model. Although
the stars are almost point-like on this scale the spatial distribution
of the non-thermal particles is anything but. The part of the
wind-wind collision region that experiences a photon flux from the
secondary star that is within a factor of 3 of the peak flux that
occurs at the apex extends to $\theta \approx 73^{\circ}$.  This
region covers 36 per cent of the sky as seen from the O-star and has
the effect of substantially reducing the change in IC emission with
the viewing angle. The presence of the WR star also reduces the level
of variability in our model.

\subsection{Effect of varying the magnetic field strength}
\begin{figure}
\includegraphics[width=8.0cm]{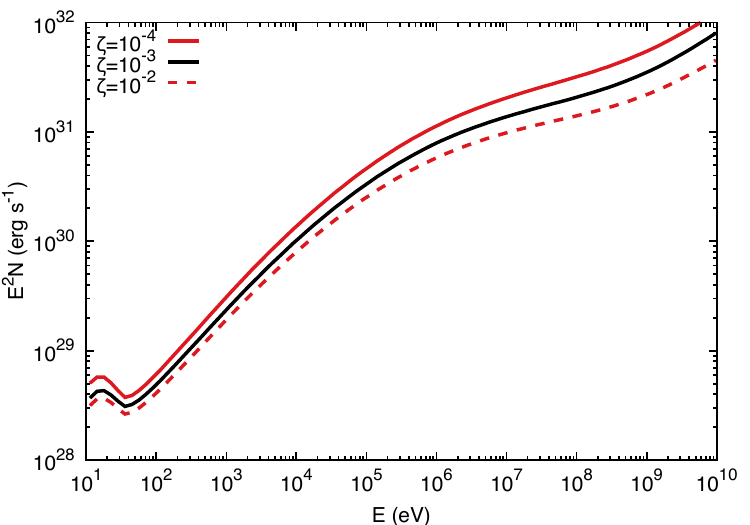}
\caption{The effect of the pre-shock magnetic flux density on the
  inverse Compton emission. The standard model has $\zeta_{\rm
  B}=10^{-3}$. $D=2\times10^{15}$\,cm and $\phi=90^{\circ}$.}
\label{fig:effectOfZetaB}
\end{figure}

In the model we are free to set the strength of the pre-shock magnetic
field. This is controlled through the value of $\zeta_{\rm B}$. Our
standard model has $\zeta_{\rm
  B}=10^{-3}$. Fig.~\ref{fig:effectOfZetaB} shows the effect
of changing $\zeta_{\rm B}$ within the range $10^{-4} \leq \zeta_{\rm
  B} \leq 10^{-2}$. Lower values of $\zeta_{\rm B}$ imply a
lower pre-shock magnetic flux density, $B_{0}$, which in turn reduces the maximum
momentum that the non-thermal protons reach ($p_{\rm max} \propto
B_{\rm 0} \propto \zeta_{\rm B}^{1/2}$). This can dramatically affect the
efficiency of the DSA, and can significantly alter the shape of the non-thermal
particle spectrum. 

\subsection{Effect of varying the injected particle momentum}
\begin{figure}
\includegraphics[width=8.0cm]{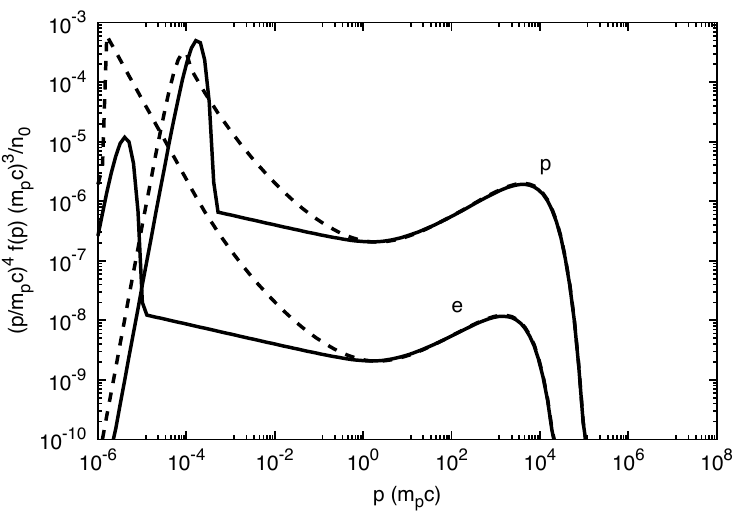}
\caption{The effect of the injected particle momentum on the
  proton and electron distributions for the on-axis WR shock as a
  function of $\chi_{\rm inj}$. The solid lines have $\chi_{\rm
    inj}=3.5$ (the standard model), while the dashed lines have $\chi_{\rm
    inj}=2.0$. $D=2\times10^{15}$\,cm.}
\label{fig:effectOfChiInjOnfppfpe}
\end{figure}

A second free parameter in the DSA model is $\chi_{\rm inj}$, which
controls the momentum of the injected particles. The effect on the
post-WR-shock non-thermal particle distributions of setting $\chi_{\rm inj} = 2.0$
is shown in Fig.~\ref{fig:effectOfChiInjOnfppfpe}. Both distributions
see an increase in the number of non-thermal particles from the thermal
peak up to momenta of order ${\rm m_{p}c}$, but show little change
above this. The variation in the proton distribution is comparable to
the differences seen when different methods are used to calculate the
DSA \citep[see, e.g., Fig.~1 in][]{Caprioli:2010}.

\begin{figure}
\includegraphics[width=8.0cm]{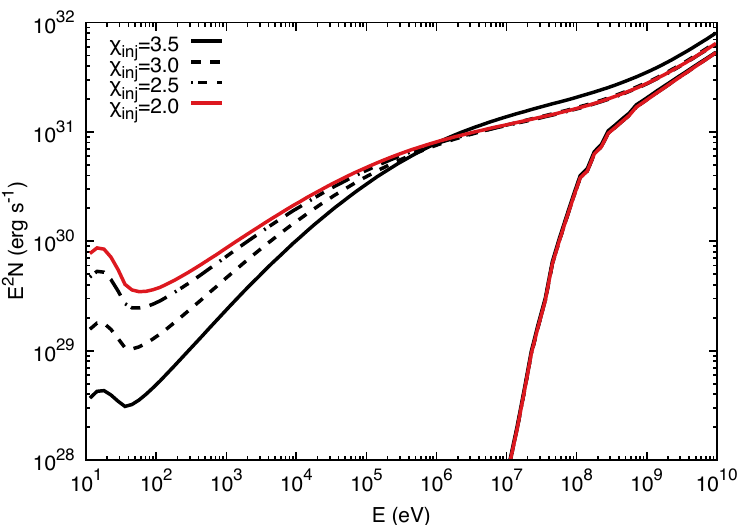}
\caption{The effect of the injected particle momentum on the
  non-thermal inverse Compton and $\pi^{0}$-decay emission. The
  standard model has $\chi_{\rm inj}=3.5$. $D=2\times10^{15}$\,cm and $\phi=90^{\circ}$.}
\label{fig:effectOfChiInj}
\end{figure}

The effect on the non-thermal emission of varying $\chi_{\rm inj}$ is shown in
Fig.~\ref{fig:effectOfChiInj}. We see that the inverse Compton
emission becomes softer as $\chi_{\rm inj}$ decreases. This is because
more electrons with $p < {\rm m_{p}c}$ (i.e with $\gamma \ltsimm
10^{3}$) take part in the DSA.
%which, for a fixed maximum momentum produces more non-thermal particles at lower energies but
%less at higher energies as the shock becomes less strongly modified \citep[see][]{Blasi:2005}.
Because the $\pi^{0}$-decay emission is produced by non-thermal
protons that exceed the threshold energy of 1.22\,GeV (see
App.~\ref{sec:appPi0}), it is not sensitive to the changes in the
non-thermal proton distribution that occur for $p < {\rm m_{p}c}$.

\subsection{Effect of varying the wind momentum ratio}
Our standard model has a wind momentum ratio $\eta=0.1$. To
examine the resulting non-thermal emission when $\eta=0.01$ we
reduce the mass-loss rate of the secondary star. This
change means that there is less energy in the winds that can ultimately be
turned into non-thermal emission. However, several effects act
together. Firstly, while less of the primary wind is shocked, a
greater fraction of the (weaker) secondary wind is shocked. Secondly,
the WCR moves closer to the secondary star. Since the wind speeds have
not changed this means that the on-axis pre- and post-shock density of
the primary and secondary wind both decline, as does the pre-shock
magnetic flux density. However, the photon flux from the secondary
star at the apex of the WCR increases. The maximum non-thermal proton
momentum at the on-axis point of the WR shock remains unchanged
($p_{\rm max}=8.5\times10^{3}\,{\rm m_{p}c}$), but reduces at the
on-axis point of the O shock to $p_{\rm max}=8.5\times10^{2}\,{\rm
  m_{p}c}$. The maximum non-thermal electron momentum at the apex of
both shocks is $p_{\rm max,e}=340\,{\rm m_{p}c}$, corresponding to
$\gamma_{\rm max,e} = 6.3\times10^{5}$ and a reduction of about a
factor of 8 from the standard model. Finally, the WCR changes shape
through a reduction in the asymptotic opening angle. 

\begin{figure}
\includegraphics[width=8.0cm]{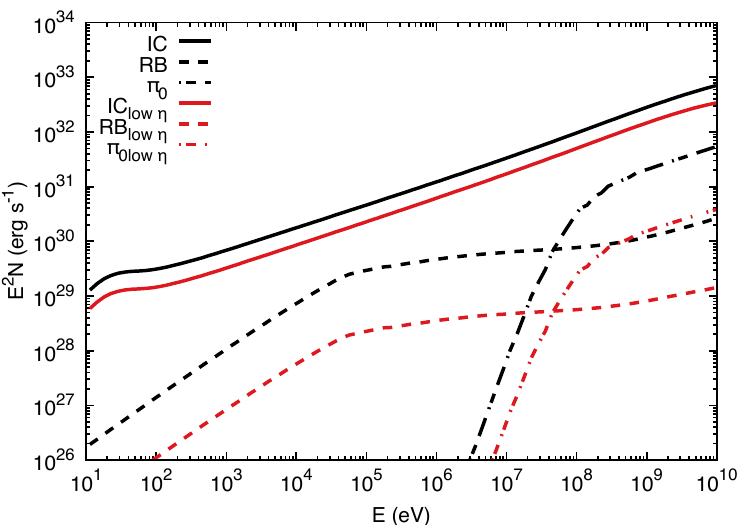}
\caption{The effect of the wind momentum ratio, $\eta$, on the
  non-thermal emission, if only adiabatic cooling of the non-thermal particles
  takes place downstream of the shock. The black lines are for
  $\eta=0.1$, and the red lines are for $\eta = 0.01$. $D=2\times10^{15}$\,cm and $\phi=90^{\circ}$.}
\label{fig:effectOfEta_uncooled}
\end{figure}

Fig.~\ref{fig:effectOfEta_uncooled} shows the effect of reducing the
wind momentum ratio, $\eta$, on the emission that would result if the
non-thermal particles were allowed only to undergo adiabatic cooling
downstream of the shocks. It shows that all three types of emission
are reduced when $\Mdot_{\rm O}$ is reduced. Thus the reduced strength
of the combined winds dominates over other factors (e.g., the enhanced
photon flux from the secondary star at the apex of the WCR). The fact
that there is less of a reduction to the inverse Compton emission than
to the relativistic bremsstrahlung and $\pi^{0}$-decay emission is
consistent with the enhanced secondary star photon flux at the WCR
somewhat offsetting the other factors noted above that act to reduce
the flux. The reduction in $\gamma_{\rm max,e}$ as $\eta$ is reduced
also affects the position of the high-energy turnover of the inverse
Compton emission (not clearly visible in
Fig.~\ref{fig:effectOfEta_uncooled}).

\begin{figure}
\includegraphics[width=8.0cm]{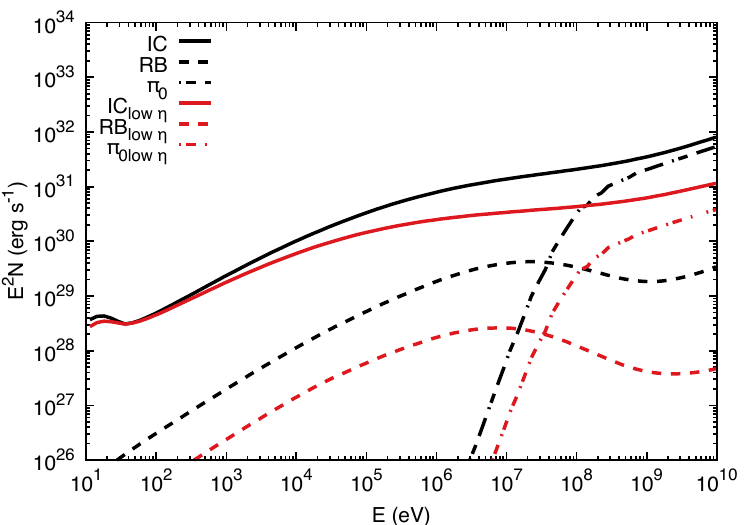}
\caption{The effect of the wind momentum ratio, $\eta$, on the
  non-thermal emission. The black lines are for
  $\eta=0.1$, and the red lines are for $\eta = 0.01$. $D=2\times10^{15}$\,cm and $\phi=90^{\circ}$.}
\label{fig:effectOfEta}
\end{figure}

Fig.~\ref{fig:effectOfEta} shows the effect on the non-thermal
emission of reducing $\eta$, with the cooling of the non-thermal
particles downstream of the shocks fully applied. The greater
reduction in the inverse Compton emission with $\eta$ compared to the
case where the downstream non-thermal particles undergo only adiabatic
cooling (see Fig.~\ref{fig:effectOfEta_uncooled}) highlights the
enhanced secondary star photon flux in this case.

\section{Summary and conclusions}
\label{sec:summary}
We have created a new model for the non-thermal emission from
colliding-wind binaries. Our model uses the \citet{Blasi:2005} model
to solve the diffusive shock acceleration of the particles at the
global shocks. We {\rm confirm earlier work} that DSA is very
efficient with our chosen parameters and assumptions, leading to
significantly modified shocks. This is the first CWB model that
self-consistently includes shock modification.

We find a complicated dependence for the scaling of the non-thermal
flux with the binary separation, $D$. If the non-thermal particles
suffer little cooling when flowing downstream from the shocks the
inverse Compton, relativistic bremsstrahlung and $\pi^{0}$-decay
emission all scale as $D^{-1}$. This occurs most readily at wide
separations and/or from leptonic emission from lower energy
particles. However, when $D$ decreases, the cooling of the non-thermal
particles increases, and simple arguments indicate that the emission
should plateau at a maximum value, becoming independent of $D$. The
$\pi^{0}$-decay emission and the lower-energy inverse Compton emission
behaves this way, but we observe more complicated behaviour for
higher-energy inverse Compton emission where the emission actually
peaks at an intermediate value of $D$ and thereafter declines as $D$
decreases further. This behaviour is caused by $p_{\rm max,e}$ also
decreasing with $D$. In real systems we may expect additional effects
caused by variations in the pre-shock wind velocities with $D$.
%For instance, the non-thermal emission should decline
%as $D$ decreases if the shock velocities reduce, which reduces the wind
%mechanical power that is available for the DSA process.

We also find that the anisotropic inverse Compton emission shows
only a moderate variation with viewing angle due to the spatial
extent of the wind-wind collision. For a system with a wind momentum
ratio of 0.1 we find that the variation with viewing angle
is limited to a factor of $\approx 3.5$. 

Reducing the wind momentum ratio from $\eta=0.1$ to $\eta=0.01$ (by
reducing the value of $\Mdot_{2}$) leads to a reduction in the
non-thermal emission due to the weaker wind-wind collision, though the
inverse Compton emission does not decline as much as the relativistic
bremsstrahlung and $\pi^{0}$-decay emission because in our model the
stellar photon flux at the apex of the WCR increases (however, in real
systems the luminosity of the secondary star may reduce too).

The first application of our new model is presented in
\citet{Mossoux:2020}, where it is compared against NuSTAR data on
Cyg\,OB2\,No.8A, a O6\,I + O5.5\,III system with a 21.9\,d period and
a slightly eccentric orbit ($e\sim0.2$). In future we will apply our
model to other particle-accelerating CWB systems, such as $\eta$\,Car,
$\gamma^{2}$\,Vel, and those in the catalogue of
\citet{DeBecker:2013}.

This is an exciting time for research into the non-thermal X-ray and
$\gamma$-ray emission from CWBs, with detections at TeV energies
expected by the future Cherenkov Telescope Array \citep[CTA;
see][]{Chernyakova:2019}.  Future improvements to our model will
include calculations of the thermal free-free and synchrotron
emission, the creation of and emission from secondary particles, and
the addition of free-free and photon-photon absorption. Radiative
inhibition \citep{Stevens:1994} and braking \citep*{Gayley:1997}, and
orbital effects \citep{Pittard:2009}, will also be examined in future
work.

\section*{Acknowledgements}
We thank the referee for some useful suggestions which improved the
quality of the paper. The calculations herein were performed on the
DiRAC 1 Facility at Leeds jointly funded by STFC, the Large Facilities
Capital Fund of BIS and the University of Leeds and on other
facilities at the University of Leeds. Data for the figures in this
paper are available from \url{https://doi.org/XXX}.  GER was supported
by grants AYA2016-76012-C3-1-P (Ministerio de Educaci\'{o}n, Cultura y
Deporte, Espa\~{n}a) and PIP 0338 (CONICET, Argentina).

%%%%%%%%%%%%%%%%%%%%%%%%%%%%%%%%%%%%%%%%%%%%%%%%%%

%%%%%%%%%%%%%%%%%%%% REFERENCES %%%%%%%%%%%%%%%%%%

% The best way to enter references is to use BibTeX:

%\bibliographystyle{mnras}
%\bibliography{example} % if your bibtex file is called example.bib

\begin{thebibliography}{99}
\bibitem[\protect\citeauthoryear{Abdo et al.}{2010}]{Abdo:2010}
Abdo A.A., et al., 2010, ApJ, 723, 649
\bibitem[\protect\citeauthoryear{Aharonian \& Atoyan}{2000}]{Aharonian:2000}
Aharonian F.A., Atoyan A.M., 2000, A\&A, 362, 937
\bibitem[\protect\citeauthoryear{Balbo \& Walter}{2017}]{Balbo:2017}
Balbo M., Walter R., 2017, A\&A, 603, A111
\bibitem[\protect\citeauthoryear{Bednarek \& Pabich}{2011}]{Bednarek:2011}
Bednarek W., Pabich J., 2011, A\&A, 530, A49
\bibitem[\protect\citeauthoryear{Benaglia}{2016}]{Benaglia:2016}
Benaglia P., 2016, PASA, 33, 17
\bibitem[\protect\citeauthoryear{Benaglia \& Romero}{2003}]{Benaglia:2003}
Benaglia P., Romero G.E., 2003, A\&A, 399, 1121
\bibitem[\protect\citeauthoryear{Benaglia et al.}{2015}]{Benaglia:2015}
Benaglia P., Marcote B., Mold\'{o}n J., Nelan E., De Becker M., Dougherty S.M., Koribalski B.S., 2015, A\&A, 579, A99
\bibitem[\protect\citeauthoryear{Berezinskii et al.}{1990}]{Berezinskii:1990}
Berezinskii V.S., Bulanov S.V., Dogiel V.A., Ptuskin V.S., 1990, ``Astrophysics of Cosmic Rays'' (North-Holland, Amsterdam)
\bibitem[\protect\citeauthoryear{Blasi, Gabici \& Vannoni}{Blasi et al.}{2005}]{Blasi:2005}
Blasi P., Gabici S., Vannoni G., 2005, MNRAS, 361, 907
\bibitem[\protect\citeauthoryear{Blomme et al.}{2017}]{Blomme:2017}
Blomme R., Fenech D.M., Prinja R.K., Pittard J.M., Morford J.C., 2017,
A\&A, 608, A69
\bibitem[\protect\citeauthoryear{Blomme et al.}{2013}]{Blomme:2013}
Blomme R., et al., 2013, A\&A, 550, A90
\bibitem[\protect\citeauthoryear{Blumenthal \& Gould}{1970}]{Blumenthal:1970}
Blumenthal G.R., Gould R.J., 1970, Rev. Mod. Phys., 42, 237
\bibitem[\protect\citeauthoryear{Brookes}{2016}]{Brookes:2016}
Brookes D.P., 2016, PhD thesis, The University of Birmingham
\bibitem[\protect\citeauthoryear{Cant\'{o}, Raga \& Wilkin}{Cant\'{o} et al.}{1996}]{Canto:1996}
Cant\'{o} J., Raga A.C., Wilkin F.P., 1996, ApJ, 469, 729 
\bibitem[\protect\citeauthoryear{Caprioli et al.}{2009}]{Caprioli:2009}
Caprioli D., Blasi P., Amato E., Vietri M., 2009, MNRAS, 395, 895
\bibitem[\protect\citeauthoryear{Caprioli et al.}{2010}]{Caprioli:2010}
Caprioli D., Kang H., Vladimirov A.E., Jones T.W., 2010, MNRAS, 407, 1773
\bibitem[\protect\citeauthoryear{Cerutti}{2007}]{Cerutti:2007}
Cerutti B., 2007, Master thesis, Ecole Nationale Sup\'{e}rieure de
Physique de Grenoble, INPG
\bibitem[\protect\citeauthoryear{Cerutti}{2010}]{Cerutti:2010}
Cerutti B., 2010, PhD thesis ``High-energy gamma-ray emission in compact
binaries'', Astrophysics, Universit\'{e} de Grenoble%, available at https://inis.iaea.org/search/search.aspx?orig_q=RN:46026940
%\bibitem[\protect\citeauthoryear{Cerutti, Dubus \& Henri}{Cerutti et al.}{2007}]{Cerutti:2007}
%Cerutti B., Dubus G., Henri G., 2007, in SF2A-2007: Proceedings of the Annual meeting of the French Society of Astronomy and Astrophysics held in Grenoble, France, July 2-6, 2007, eds. J.~Bouvier, A.~Chalabaev, and C.~Charbonnel, p158
\bibitem[\protect\citeauthoryear{Chernyakova et al.}{2019}]{Chernyakova:2019}
Chernyakova M., et al., 2019, A\&A, 631, A177
\bibitem[\protect\citeauthoryear{Corcoran}{2005}]{Corcoran:2005}
Corcoran M.F., 2005, AJ, 129, 2018
\bibitem[\protect\citeauthoryear{Corcoran et al.}{2010}]{Corcoran:2010}
Corcoran M.F., Hamaguchi K., Pittard J.M., Russell C.M.P., Owocki
S.P., Parkin E.R., Okazaki A., 2010, ApJ, 725, 1528
\bibitem[\protect\citeauthoryear{Damineli et al.}{2008}]{Damineli:2008}
Damineli A., et al., 2008, MNRAS, 384, 1649
\bibitem[\protect\citeauthoryear{De Becker \& Raucq}{2013}]{DeBecker:2013}
De Becker M., Raucq F., 2013, A\&A, 558, A28
\bibitem[\protect\citeauthoryear{del Palacio et al.}{2016}]{delPalacio:2016}
del Palacio S., Bosch-Ramon V., Romero G.E., Benaglia P., 2016, A\&A, 591, 139
\bibitem[\protect\citeauthoryear{Dougherty et al.}{2005}]{Dougherty:2005}
Dougherty S.M., Beasley A.J., Claussen M.J., Zauderer B.A., Bolingbroke N.J., 2005, ApJ, 623, 447
\bibitem[\protect\citeauthoryear{Dougherty et al.}{2003}]{Dougherty:2003}
Dougherty S.M., Pittard J.M., Kasian L., Coker R.F., Williams P.M.,
Lloyd H.M., 2003, A\&A, 409, 217
\bibitem[\protect\citeauthoryear{Dougherty \& Pittard}{2006}]{Dougherty:2006}
Dougherty S.M., Pittard J.M., 2006, in Proc. of the 8th European VLBI
Network Symp., 49
\bibitem[\protect\citeauthoryear{Dougherty, Williams \&
    Pollacco}{Dougherty et al.}{2000}]{Dougherty:2000}
Dougherty S.M., Williams P.M., Pollacco D.L., 2000, MNRAS, 316, 143
\bibitem[\protect\citeauthoryear{Dubus, Cerutti \& Henri}{Dubus et al.}{2008}]{Dubus:2008}
Dubus G., Cerutti B., Henri G., 2008, A\&A, 477, 691
\bibitem[\protect\citeauthoryear{Dzib et al.}{2013}]{Dzib:2013}
Dzib S.A., Rodr\'{i}guez L.F., Loinard L., Mioduszewski A.J.,
Ortiz-Le\'{o}n G.N., Araudo A.T., 2013, ApJ, 763, 139
\bibitem[\protect\citeauthoryear{Eichler \& Usov}{1993}]{Eichler:1993}
Eichler D., Usov V., 1993, ApJ, 402, 271
\bibitem[\protect\citeauthoryear{Farnier, Walter \& Leyder}{Farnier et
    al.}{2011}]{Farnier:2011}
Farnier C., Walter R., Leyder J.-C., 2011, A\&A, 526, A57
\bibitem[\protect\citeauthoryear{Gayley, Owocki \& Cranmer}{Gayley et al.}{1997}]{Gayley:1997}
Gayley K.G., Owocki S.P., Cranmer S.R., 1997, ApJ, 475, 786
\bibitem[\protect\citeauthoryear{Ginzburg \& Syrovatskii}{1964}]{Ginzburg:1964}
Ginzburg V., Syrovatskii S., 1964, ``The Origin of Cosmic Rays'' (New York: Macmillan)
\bibitem[\protect\citeauthoryear{Grimaldo et al.}{2019}]{Grimaldo:2019}
Grimaldo E., Reimer A., Kissmann R., Niederwanger F., Reitberger K.,
2019, ApJ, 871, 55
\bibitem[\protect\citeauthoryear{Hamaguchi et al.}{2007}]{Hamaguchi:2007}
Hamaguchi K., et al., 2007, ApJ, 663, 522
\bibitem[\protect\citeauthoryear{Hamaguchi et al.}{2014}]{Hamaguchi:2014}
Hamaguchi K., et al., 2014, ApJ, 795, 119
\bibitem[\protect\citeauthoryear{Hamaguchi et al.}{2018}]{Hamaguchi:2018}
Hamaguchi K., et al., 2018, Nature Astronomy, 2, 731
\bibitem[\protect\citeauthoryear{Heitler}{1954}]{Heitler:1954}
Heitler W., 1954, ``Quantum Theory of Radiation''
\bibitem[\protect\citeauthoryear{H.E.S.S. collaboration}{2020}]{HESS:2020}
H.E.S.S. Collaboration, Abdalla et al., 2020, arXiv:2002.02336
\bibitem[\protect\citeauthoryear{Kelner, Aharonian \& Bugayov}{Kelner et al.}{2006}]{Kelner:2006}
Kelner S.R., Aharonian F.A., Bugayov V.V., 2006, Phys. Rev. D, 74, 034018
\bibitem[\protect\citeauthoryear{Khangulyan \& Aharonian}{2005}]{Khangulyan:2005}
Khangulyan D., Aharonian F.A., 2005, in American Institute of Physics
Conference Series, 745, High Energy Gamma-Ray Astronomy,
ed. F.A. Aharonian, H.J. V\"{o}lk, \& D. Horns, 359
%\bibitem[\protect\citeauthoryear{Leser et al.}{2017}]{Leser:2017}
%Leser E., et al., 2017, Proceedings of Science (ICRC2017), 717
\bibitem[\protect\citeauthoryear{Leyder, Walter \& Rauw}{Leyder et al.}{2008}]{Leyder:2008}
Leyder J.-C., Walter R., Rauw G., 2008, A\&A, 477, L29
\bibitem[\protect\citeauthoryear{Leyder, Walter \& Rauw}{Leyder et al.}{2010}]{Leyder:2010}
Leyder J.-C., Walter R., Rauw G., 2010, A\&A, 524, A59
\bibitem[\protect\citeauthoryear{Madura et al.}{2013}]{Madura:2013}
Madura T.I., et al., 2013, MNRAS, 436, 3820
\bibitem[\protect\citeauthoryear{Manolakou, Horns \& Kirk}{Manolakou et al.}{2007}]{Manolakou:2007}
Manolakou K., Horns D., Kirk J.G., 2007, A\&A, 474, 689
\bibitem[\protect\citeauthoryear{Mehner et al.}{2010}]{Mehner:2010}
Mehner A., Davidson K., Ferland G.J., Humphreys R.M., 2010, ApJ, 710, 729
\bibitem[\protect\citeauthoryear{Mossoux et al.}{2020}]{Mossoux:2020}
Mossoux E., Pittard J.M., Rauw G., Naz\'{e} Y., 2020, A\&A, accepted (arXiv:2003.10262)
\bibitem[\protect\citeauthoryear{O'Connor et al.}{2005}]{O'Connor:2005}
O'Connor E.P., Dougherty S.M., Pittard J.M., Williams P.M., 2005, in
Massive Stars and High-Energy Emission in OB Associations, eds. G. Rauw,
Y. Naz\'{e}, R. Blomme, \& E. Gosset, 81
\bibitem[\protect\citeauthoryear{Ohm et al.}{2015}]{Ohm:2015}
Ohm S., Zabalza V., Hinton J.A., Parkin E.R., 2015, MNRAS, 449, L132
\bibitem[\protect\citeauthoryear{Okazaki et al.}{2008}]{Okazaki:2008}
Okazaki A.T., Owocki S.P., Russell C.M.P., Corcoran M.F., 2008, MNRAS,
388, L39
\bibitem[\protect\citeauthoryear{Ortiz-Le\'{o}n et al.}{2011}]{Ortiz-Leon:2011}
Ortiz-Le\'{o}n G., Loinard L., Rodr\'{i}guez L.F., Mioduszewski A.J.,
Dzib S.A., 2011, ApJ, 737, 30
\bibitem[\protect\citeauthoryear{Parkin \& Pittard}{2008}]{Parkin:2008}
Parkin E.R., Pittard J.M., 2008, MNRAS, 388, 1047
\bibitem[\protect\citeauthoryear{Parkin et al.}{2009}]{Parkin:2009}
Parkin E.R., Pittard J.M., Corcoran M.F., Hamaguchi K., Stevens I.R., 2009, MNRAS, 394, 1758
\bibitem[\protect\citeauthoryear{Parkin et al.}{2011}]{Parkin:2011}
Parkin E.R., Pittard J.M., Corcoran M.F., Hamaguchi K., 2011, ApJ, 726, 105
\bibitem[\protect\citeauthoryear{Pittard}{2009}]{Pittard:2009}
Pittard J.M., 2009, MNRAS, 396, 1743
\bibitem[\protect\citeauthoryear{Pittard \& Corcoran}{2002}]{Pittard:2002}
Pittard J.M., Corcoran M.F., 2002, A\&A, 383, 636
\bibitem[\protect\citeauthoryear{Pittard \& Dawson}{2018}]{Pittard:2018}
Pittard J.M., Dawson B., 2018, MNRAS, 477, 5640
\bibitem[\protect\citeauthoryear{Pittard \& Dougherty}{2006}]{Pittard:2006b}
Pittard J.M., Dougherty S.M., 2006, MNRAS, 372, 801
\bibitem[\protect\citeauthoryear{Pittard et al.}{2006}]{Pittard:2006a}
Pittard J.M., Dougherty S.M., Coker R.F., O'Connor E., Bolingbroke
N.J., 2006, A\&A, 446, 1001
\bibitem[\protect\citeauthoryear{Pshirkov}{2016}]{Pshirkov:2016}
Pshirkov M.S., 2016, MNRAS, 457, L99
\bibitem[\protect\citeauthoryear{Reimer, Pohl \& Reimer}{Reimer et al.}{2006}]{Reimer:2006}
Reimer A., Pohl M., Reimer O., 2006, ApJ, 644, 1118
\bibitem[\protect\citeauthoryear{Reitberger et al.}{2014a}]{Reitberger:2014a}
Reitberger K., Kissmann R., Reimer A., Reimer O., Dubus G., 2014a, ApJ, 782, 96
\bibitem[\protect\citeauthoryear{Reitberger et al.}{2014b}]{Reitberger:2014b}
Reitberger K., Kissmann R., Reimer A., Reimer O., 2014b, ApJ, 789, 87
\bibitem[\protect\citeauthoryear{Reitberger et al.}{2015}]{Reitberger:2015}
Reitberger K., Reimer A., Reimer O., Takahashi H., 2015, A\&A, 577, A100
\bibitem[\protect\citeauthoryear{Reitberger et al.}{2017}]{Reitberger:2017}
Reitberger K., Kissmann R., Reimer A., Reimer O., 2017, ApJ, 847, 40
\bibitem[\protect\citeauthoryear{Reitberger et al.}{2012}]{Reitberger:2012}
Reitberger K., Reimer O., Reimer A., Werner M., Egberts K., Takahashi H., 2012, A\&A, 544, A98
\bibitem[\protect\citeauthoryear{Rybicki \& Lightman}{1979}]{Rybicki:1979}
Rybicki G.B., Lightman A.P., 1979, ``Radiative Processes in Astrophysics''
\bibitem[\protect\citeauthoryear{Sekiguchi et al.}{2009}]{Sekiguchi:2009}
Sekiguchi A., Tsujimoto M., Kitamoto S., Ishida M., Hamaguchi K., Mori
H., Tsuboi Y., 2009, PASJ, 61, 629
\bibitem[\protect\citeauthoryear{Stecker}{1971}]{Stecker:1971}
Stecker F.W., 1971, Cosmic Gamma Rays, Baltimore: Mono Book corp.
\bibitem[\protect\citeauthoryear{Stevens, Blondin \& Pollock}{Stevens
    et al.}{1992}]{Stevens:1992}
Stevens I.R., Blondin J.M., Pollock A.M.T., 1992, ApJ, 386, 265
\bibitem[\protect\citeauthoryear{Stevens \& Pollock}{1994}]{Stevens:1994}
Stevens I.R., Pollock A.M.T., 1994, MNRAS, 269, 226 
\bibitem[\protect\citeauthoryear{Tavani et al.}{2009}]{Tavani:2009}
Tavani M., et al., 2009, ApJL, 698, L142
\bibitem[\protect\citeauthoryear{Vila}{2012}]{Vila:2012}
Vila G.S., 2012, PhD thesis ``Radiative models for jets in X-ray binaries'', Universidad de Buenos Aires
\bibitem[\protect\citeauthoryear{Williams et al.}{1997}]{Williams:1997}
Williams P.M., Dougherty S.M., Davis R.J., van der Hucht K.A., Bode M.F., Setia Gunawan D.Y.A., 1997, MNRAS, 289, 10
\bibitem[\protect\citeauthoryear{Zdziarski}{1989}]{Zdziarski:1989}
Zdziarski A.A., 1989, ApJ, 342, 1108
\end{thebibliography}

% Alternatively you could enter them by hand, like this:
% This method is tedious and prone to error if you have lots of references

%%%%%%%%%%%%%%%%%%%%%%%%%%%%%%%%%%%%%%%%%%%%%%%%%%

%%%%%%%%%%%%%%%%% APPENDICES %%%%%%%%%%%%%%%%%%%%%

\appendix

\section{Cooling of the  non-thermal particles}
\label{sec:AppCooling}
In this appendix we provide equations for the cooling rate of
non-thermal electrons and protons.

The cooling rate of electrons is given by \citep[cf.][]{Ginzburg:1964,Manolakou:2007}
\begin{equation}
\label{eq:electronGammaDot}
\frac{d \gamma_{\rm e}}{dt} = b_{\rm S}\gamma_{\rm e}^{2} + b_{\rm
  IC}\gamma_{\rm e}^{2}F_{\rm KN}(\gamma_{\rm e}) + b_{\rm C}({\rm ln}\gamma_{\rm e} + b_{\rm
  C}^{0}) + b_{\rm B}\gamma_{\rm e}({\rm ln}\gamma_{\rm e} + b_{\rm
  B}^{0}) + \frac{v\gamma_{\rm e}}{R},
\end{equation}
where the coefficients $b_{\rm S}$, $b_{\rm IC}$, $b_{\rm C}$ and
$b_{\rm B}$ for the synchrotron, inverse Compton, coulombic and
bremsstrahlung losses are given by
\begin{equation}
b_{\rm S} = \frac{4 \sigma_{\rm T}}{3 m_{\rm e}c} U_{\rm B} =
1.292\times10^{-15}(B/{\rm mG})^{2}\, {\rm s^{-1}},
\end{equation}
\begin{equation}
b_{\rm IC} = b_{\rm S}\frac{U_{\rm ph}}{U_{\rm B}} =
5.204\times10^{-20} (U_{\rm ph}/{\rm eV\,cm^{-3}})\, {\rm s^{-1}},
\end{equation}
\begin{equation}
b_{\rm C} = \frac{2 \pi e^{4} n_{\rm e}}{m_{\rm e}^{2}c^{3}} =
1.491\times10^{-14}n_{\rm e}\, {\rm s^{-1}},
\end{equation}
and
\begin{equation}
b_{\rm B} = \frac{4 e^{6} n_{\rm e}}{m_{\rm e}^{2}c^{4}\hbar} =
1.37\times10^{-16}n_{\rm e}\, {\rm s^{-1}}.
\end{equation}
The constants $b_{\rm C}^{0}$ and $b_{\rm B}^{0}$ are given by
\begin{equation}
b_{\rm C}^{0} = {\rm ln}\left(\frac{m_{\rm e}^{3}c^{4}}{4 e^{2} n_{\rm
      e} \hbar^{2}}\right) + \frac{3}{4} = -{\rm ln}n_{\rm e} + 73.4,
\end{equation}
and
\begin{equation}
b_{\rm B}^{0} = {\rm ln}2 - \frac{1}{3} = 0.36.
\end{equation}
In these equations, $\sigma_{\rm T}$ is the Thomson cross section,
$U_{\rm ph}$ and $U_{\rm B}$ are the photon and magnetic field energy
densities, respectively, $n_{\rm e}$ is the electron number density, while $c$, $m_{\rm e}$ and $e$ are the speed
of light, and the electron mass and charge. For a black-body
distribution of target photons, 
\begin{equation}
F_{\rm KN}(\gamma_{\rm e}) \approx (1 + 4\gamma_{\rm e}\epsilon_{\rm eff})^{-3/2},
\end{equation}
where
\begin{equation}
\epsilon_{\rm eff} = 2.8\frac{kT}{m_{\rm e}c^{2}}.
\end{equation}
This approximation takes into account the full Klein-Nishina
cross-section for Compton scattering, and is valid for an anisotropic
target photon field provided the electron distribution is isotropic
\citep[see][and references therein]{Manolakou:2007}.

The last term in Eq.~\ref{eq:electronGammaDot} is due to the adiabatic
cooling. Here $v$ is the flow speed along the contact discontinuity
and $R$ is the distance of the gas from its star. We assume that the
hot plasma expands almost spherically as it flows out of the system,
consistent with the approach taken by \citet{delPalacio:2016}. 

In addition to cooling the non-thermal particles, adiabatic expansion also
reduces their number density. We assume
that a change in volume occurs when the non-thermal particles flow from one
segment to the next, and that this change is related to the difference in
the immediate post-shock density of the thermal plasma between the segments.
Specifically, we assume that $\rho_{1} V_{1} = \rho_2 V_2$, where
$\rho_{1(2)}$ and $V_{1(2)}$ are the density of the thermal particles and volume of the
non-thermal particles in segment 1(2). The change in volume,
$dV = V_2 - V_1$. Thus $dV/V_2 = (\rho_1/\rho_2 - 1)$.

The change of the electron Lorentz factor with time in
  Eq.~\ref{eq:electronGammaDot} is defined to be positive for electron
  cooling, so
\begin{equation}
\dot{E_{\rm e}} = -m_{\rm e}c^{2}\frac{d \gamma_{\rm e}}{dt}.
\end{equation}

The cooling rate of non-thermal protons is given by
\begin{equation}
\label{eq:proton_cooling}
\frac{d \gamma_{\rm p}}{dt} = cn_{\rm p}\gamma_{\rm p}K_{\rm pp}\sigma_{\rm
  pp}(\gamma_{\rm p}) + \frac{v\gamma_{\rm p}}{R},
\end{equation}
where $n_{\rm p}$ is the number density of thermal protons,
$\sigma_{\rm pp}$ is the total inelastic cross section and
$K_{\rm pp} \approx 0.5$ is the total inelasticity of the
interaction. $\sigma_{\rm pp}$ can be approximated as \citep{Kelner:2006}
\begin{equation}
\sigma_{\rm pp}(E_{\rm p}) = (34.3 + 1.88L + 0.25L^{2})\left[1 -
  \left(\frac{E_{\rm th}}{E_{\rm p}}\right)^{4}\right]^{2}\,{\rm mb},
\end{equation}
where $E_{\rm th} = 1.22$\,GeV is the threshold energy for the
production of a single $\pi^{0}$.

\section{Emissivities}
\label{sec:appA}
In this appendix we provide equations for the emissivity calculations
in our models (see \citet{Cerutti:2007} and \citet{Vila:2012} for further details). 

\begin{figure*}
\includegraphics[width=16.0cm]{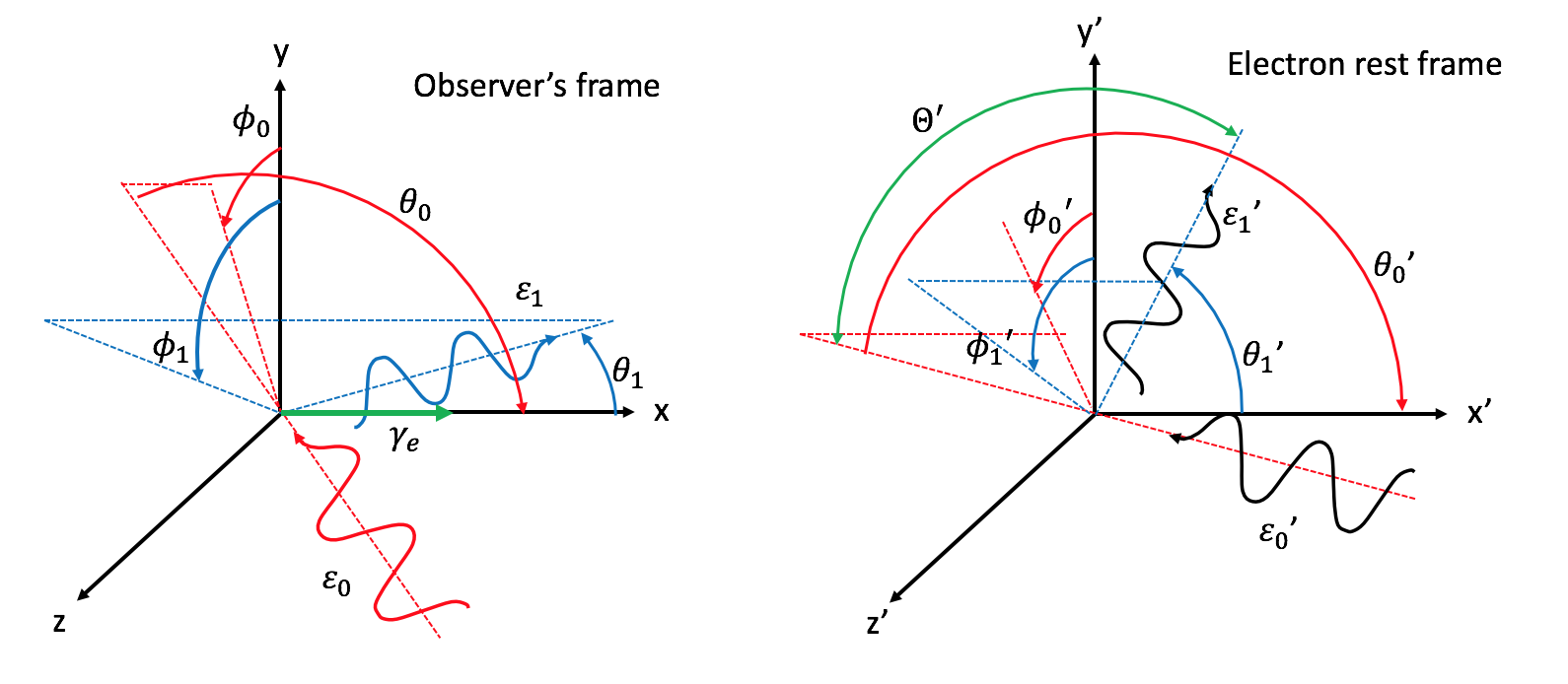}
\caption{The inverse Compton scattering geometry and parameters in the
  observer's frame (left) and the rest frame of the electron
  (right). The incident and scattered photons are represented by waves
  and the green arrow shows the direction of motion of the electron in
  the lab frame. The Lorentz boost from the observer to the rest frame
  of the electron is along the $x$-axis.}
\label{fig:ICframes}
\end{figure*}

\subsection{Anisotropic inverse Compton emission}
\label{sec:appIC}
Consider a target photon scattering off an electron that is moving
with velocity $v=\beta\,c$. One can define two reference frames: K is
the observer (lab) frame and K' is the rest frame of the electron. In
the lab frame the angle between the incident photon and the electron
is $\theta_{0}$, and the photon energy is $\epsilon_{0}$. All the
quantities defined in K that are measured in K' are written with a
``prime''. Thus in the rest frame of the electron, the angle and
energy of the incident photon are $\theta_{0}'$ and $\epsilon_{0}'$,
respectively. The scattered photon moves at an angle $\theta_{1}$ from
the direction vector of the electron in the lab frame, and at an angle
of $\theta_{1}'$ in the electron rest frame. The scattered photon has
an energy $\epsilon_{1}$ in the lab frame, and an energy
$\epsilon_{1}'$ in the electron rest frame. In the lab frame the
incident photon has an azimuthal angle $\phi_{0}$, while the scattered
photon has an azimuthal angle $\phi_{1}$. These angles are
respectively $\phi_{0}'$ and $\phi_{1}'$ in the electron rest
frame. The geometry and parameters are shown in
Fig.~\ref{fig:ICframes}. The derivation below closely follows the work
in \citet{Cerutti:2007}.

The Compton formula gives
\begin{equation}
\epsilon_{1}' = \frac{\epsilon_{0}'}{1 + \frac{\epsilon_{0}'}{m_{\rm e}
    c^{2}}(1 - \cos\Theta')},
\end{equation}
where the scattering angle $\Theta'$ is a function of the other angles
of the problem:
\begin{equation}
\cos\Theta' = \cos\theta_{1}'\cos\theta_{0}' + \sin\theta_{1}'\sin\theta_{0}'\cos(\phi_{1}'-\phi_{0}').
\end{equation}

%In the relativistic regime, Lorentz transformations between the two
%frames link the two energies via the relativistic parameters
%$\beta=v/c$ and $\gamma$.

The differential
cross section per solid angle $d\Omega_{1}'$ and per energy
$\epsilon_{1}'$ of the Compton scattering for unpolarized radiation is
given by the Klein-Nishina formula \citep[see, e.g.,][]{Heitler:1954,Rybicki:1979}
\begin{multline}
\label{eq:KNxsection}
%This is Eq 3.6 in Cerutti:2007
\frac{d\sigma}{d\epsilon_{1}'d\Omega_{1}'} = \frac{r_{\rm e}^{2}}{2}
\left(\frac{\epsilon_{1}'}{\epsilon_{0}'}\right)^{2}
\left(\frac{\epsilon_{1}'}{\epsilon_{0}'} +
  \frac{\epsilon_{0}'}{\epsilon_{1}'} - \sin^{2}\Theta'\right) \times \\
\delta\left(\epsilon_{1}' - \frac{\epsilon_{0}'}{1 +
    \frac{\epsilon_{0}'}{m_{\rm e} c^{2}}(1 - \cos\Theta')}\right), 
\end{multline}
where $r_{\rm e}$ is the classical electron radius.

Now consider a monoenergetic and unidimensional photon distribution
interacting with a single energetic electron of energy
$E_{\rm e}=\gamma m_{\rm e}c^{2}$. In the observer's (lab) frame
this distribution (in units of ${\rm photons/cm^{3}/erg/sr}$) can be
written as 
\begin{equation}
\label{eq:incomingPhotonDistribution}
n_{\rm ph} = \frac{dn}{d\epsilon d\Omega} = n_{0}\delta(\epsilon-\epsilon_{0})\delta(\theta-\theta_{0})\delta(\phi-\phi_{0}),
\end{equation}
where $\epsilon$ is the energy of the incident photons, and $\theta$
and $\phi$ are the polar and azimuthal angle \citep[see Figure~3.1
in][]{Cerutti:2007}. The polar axis $x$ is parallel to the initial
electron momentum, so that the polar angle $\theta_{0}$ is also the
collision angle.

Since $\frac{dn}{d\epsilon}$ is a Lorentz invariant,
  $\frac{dn}{d\epsilon}=\frac{dn'}{d\epsilon'}$, so
\begin{equation}
\frac{dn'}{d\epsilon'd\Omega'} = \frac{dn}{d\epsilon
  d\Omega}\frac{d\Omega}{d\Omega'}.
\end{equation}
Using the Doppler shift formulae,
Eq.~\ref{eq:incomingPhotonDistribution} becomes
\begin{multline}
%before Cerutti:2007 Eq 3.11
\frac{dn}{d\epsilon d\Omega} = n_{0}\delta(\epsilon'\gamma(1+\beta
\cos\theta')-\epsilon_{0})\delta\left(\frac{\cos \theta' + \beta}{1 +
    \beta\cos\theta'}-\cos\theta_{0}\right)\times \\ \delta(\phi'-\phi_{0}'),
\end{multline}
and 
\begin{equation}
\frac{d\Omega}{d\Omega'} = \gamma^{2}(1 - \beta\cos\theta)^{2} =
\frac{1}{\gamma^{2}(1 + \beta\cos\theta')^{2}}.
\end{equation}

The Dirac distribution has the property that for a function $f(x)$
where for all $i$, $f(x_{i})=0$, then
\begin{equation}
\label{eq:dirac_distribution_property}
\delta(f(x)) = \sum_{i} \frac{1}{|df/dx|_{x=x_{i}}}\delta(x-x_{i}).
\end{equation}
It is then possible to express the differential photon density in K'
as
\begin{multline}
%Eq 3.12 Cerutti:2007
\frac{dn'}{d\epsilon'd\Omega'} = n_{0}\gamma(1 -
\beta\cos\theta_{0})\delta(\epsilon' - \epsilon_{0}\gamma(1 -
\beta\cos\theta_{0})) \times \\ \delta\left(\cos\theta' -
\frac{\cos\theta_{0}-\beta}{1 - \beta\cos\theta_{0}}\right)\delta(\phi'-\phi_{0}').
\end{multline}

\begin{figure*}
\includegraphics[width=16.0cm]{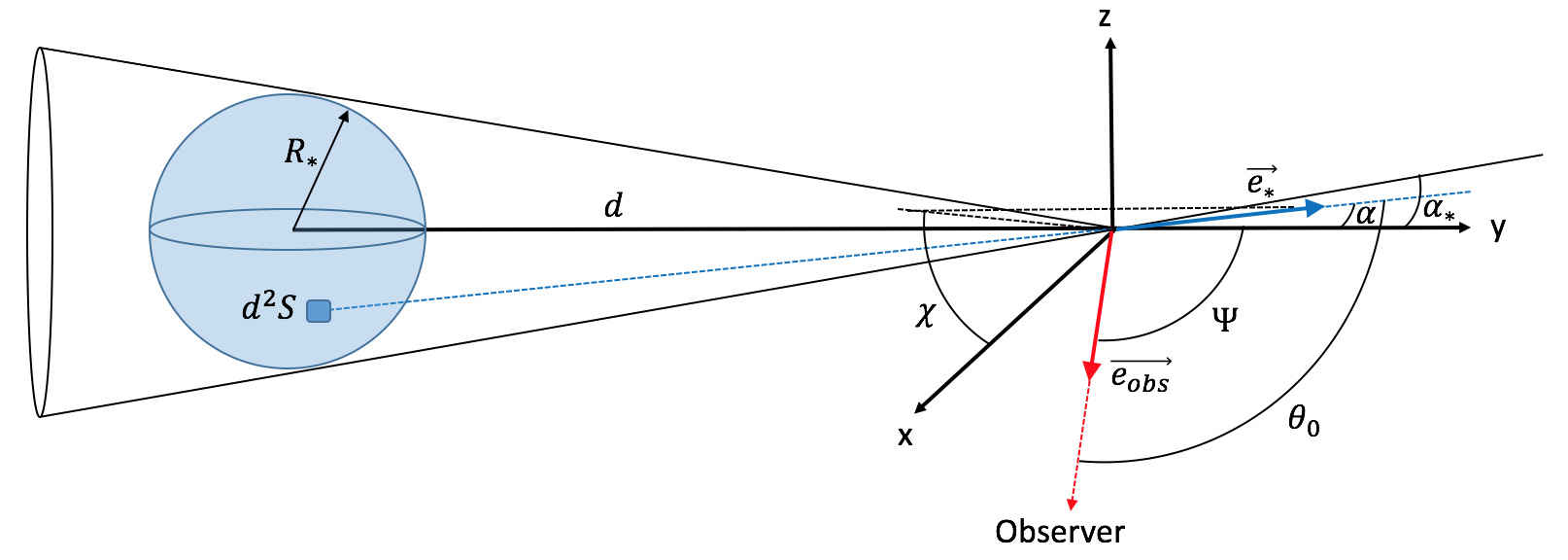}
\caption{The geometry of the star-electron-observer. The blue arrow
  shows the direction of the incoming stellar photon, and the red
  arrow shows the direction of the up-scattered photon. Because the
  inverse Compton emission from energetic electrons is highly beamed
  this is also the direction vector of the electron. The aperture
  angle of the star is $\alpha_{*}$, the viewing angle is $\psi$, and
  the scattering angle is $\theta_{0}$. The $y$-axis is defined to be
  the polar axis. $\chi$ is the azimuthal angle in the $xz$-plane, and
  $\psi$ is in the $xy$-plane.}
\label{fig:starElectronObserverGeometry}
\end{figure*}

To obtain the inverse Compton spectrum per electron, we first need to
determine the differential number of scattered photons (photons/s/sr$^{2}$/erg$^{2}$) in
the rest frame of the electron, which is
\begin{equation}
%Eq 3.13 in Cerutti:2007
\frac{dN}{dt'd\epsilon_{1}'d\Omega_{1}'d\epsilon'd\Omega'} =
\frac{dn'}{d\epsilon'd\Omega'} c \frac{d\sigma}{d\epsilon_{1}'d\Omega_{1}'}.
\end{equation}
However, the observer is interested in the differential number of
scattered photons per electron, per unit of time, per unit of energy
$\epsilon_{1}$ and per unit of solid angle $\Omega_{1}$ in the lab
frame, which is given by
\begin{equation}
%Eq 3.14 in Cerutti:2007
\frac{dN}{dt d\epsilon_{1}d\Omega_{1}} =
\int_{\Omega'}\int_{\epsilon'}
\frac{dN}{dt'd\epsilon_{1}'d\Omega_{1}'d\epsilon'd\Omega'}
\frac{dt'}{dt} \frac{d\Omega_{1}'}{d\Omega_{1}}
\frac{\epsilon_{1}'}{\epsilon_{1}} d\Omega'd\epsilon'.
\end{equation}
The Jacobian of the Lorentz transformation from K' to K is
\begin{equation}
%Eq 3.15 in Cerutti:2007
\frac{dt'}{dt} \frac{d\Omega_{1}'}{d\Omega_{1}}
\frac{\epsilon_{1}'}{\epsilon_{1}} = \frac{1}{\gamma}
\frac{1}{\gamma^{2}(1 - \beta\cos\theta_{1})^{2}} \gamma(1 -
\beta\cos\theta_{1}).
% = \frac{1}{\gamma^{2}(1 - \beta\cos\theta_{1})}.
\end{equation}
Thus one obtains
\begin{equation}
%Eq 3.16 in Cerutti:2007
\label{eq:cerutti2007_3.16}
\frac{dN}{dt d\epsilon_{1}d\Omega_{1}} = \frac{1}{\gamma^{2}(1 - \beta\cos\theta_{1})}\int_{\Omega'}\int_{\epsilon'}\frac{dn'}{d\epsilon'd\Omega'} c \frac{d\sigma}{d\epsilon_{1}'d\Omega_{1}'}d\Omega'd\epsilon'.
\end{equation}
Using the Dirac distribution property
(Eq.~\ref{eq:dirac_distribution_property}), and defining $\mu =
\cos\Theta'$, Eq.~\ref{eq:KNxsection} becomes
\begin{multline}
%Eq 3.27 in Cerutti:2007
\delta\left(\epsilon_{1}' - \frac{\epsilon'}{1 +
    \frac{\epsilon'}{m_{\rm e}c^{2}}(1 - \mu)}\right) = \\ \frac{1}{\left[1 -
    \frac{\epsilon_{1}'}{m_{\rm e}c^{2}}(1-\mu)\right]^{2}}
\delta\left(\epsilon' - \frac{\epsilon_{1}'}{1 -
    \frac{\epsilon_{1}'}{m_{\rm e}c^{2}}(1-\mu)}\right).
\end{multline}
Eq.~\ref{eq:cerutti2007_3.16} then becomes
\begin{multline}
%Eq 3.28 in Cerutti:2007
\frac{dN}{dt d\epsilon_{1}d\Omega_{1}} = \frac{1}{\gamma^{2}(1 - \beta\cos\theta_{1})}\int_{\Omega'}\int_{\epsilon'}n_{0}\gamma(1 -
\beta\cos\theta_{0})\times \\ \hspace{15mm} \delta(\epsilon' - \epsilon_{0}\gamma(1 -
\beta\cos\theta_{0})) \delta(\cos\theta' -
\cos\theta_{0}')\delta(\phi'-\phi_{0}') \times \\ \hspace{15mm}
\frac{r_{\rm e}^{2}}{2}\left(\frac{\epsilon_{1}'}{\epsilon'}\right)^{2}\left(\frac{\epsilon_{1}'}{\epsilon'}
  + \frac{\epsilon'}{\epsilon_{1}'} - 1 + \mu^{2}\right) \frac{1}{\left[1 -
    \frac{\epsilon_{1}'}{m_{\rm e}c^{2}}(1-\mu)\right]^{2}}\times \\ \hspace{15mm}
\delta\left(\epsilon' - \frac{\epsilon_{1}'}{1 -
    \frac{\epsilon_{1}'}{m_{\rm e}c^{2}}(1-\mu)}\right)d\epsilon'd\Omega'.
\end{multline}
These integrations are straightforward and give
\begin{multline}
%Eq 3.29 in Cerutti:2007
\frac{dN}{dtd\epsilon_{1}d\Omega_{1}} =
\frac{r_{\rm e}^{2} n_{0} c (1 - \beta \cos
  \theta_{0})}{2\gamma(1 - \beta \cos\theta_{1})} \times \\
\hspace{15mm} \left[1 + \mu^{2} +
  \left(\frac{\gamma \epsilon_{1}}{m_{\rm e} c^{2}}\right)^{2} \frac{(1 -
    \beta \cos\theta_{1})^{2}(1 - \mu)^{2}}{1 - \frac{\gamma
      \epsilon_{1}}{m_{\rm e} c^{2}}(1 - \beta
    \cos\theta_{1})(1-\mu)}\right] \times \\ \hspace{15mm}
\delta\left(\frac{\gamma \epsilon_{1}(1-\beta\cos\theta_{1})}{1 -
    \frac{\gamma \epsilon_{1}}{m_{\rm e} c^{2}}(1 -
    \beta\cos\theta_{1})(1-\mu)} - \gamma\epsilon_{0}(1 - \beta\cos\theta_{0})\right).
\end{multline}
The integration over $\Omega_{1}$ is complicated, but can be obtained
by making use of the approximation that for $\gamma >> 1$,
\begin{equation}
\mu \approx \frac{\cos \theta_{1} - \beta}{1 - \beta \cos \theta_{1}} C_{\theta_{0}},
\end{equation}
where
\begin{equation}
C_{\theta_{0}} = \frac{\cos \theta_{0} - \beta}{1 - \beta \cos \theta_{0}}.
\end{equation}
Because of this approximation, the spectrum is independent of the
azimuthal angle and the integration over the azimuthal angle
$\phi_{1}$ just multiplies it by $2\pi$. The remaining integration
over $x = \cos\theta_{1}$ is simple as well. If $\mu_{0} = \mu(x_{0})$,
the number of photons with final energy
$\epsilon_{1}$ scattered into all outward directions per unit time (${\rm photons/s/erg}$) is
then
\begin{multline}
\label{eq:anisotropicICemissivity_singleElectron}
%This is Eq.3.33 in Cerutti:2007
\frac{dN}{dtd\epsilon_{1}}(E_{\rm e},\epsilon_{0},\epsilon_{1}) =
\pi r_{\rm e}^{2} n_{0} c K \frac{(1 - \beta \cos
  \theta_{0})}{\gamma(1 - \beta x_{0})} \times \\ \left[1 + \mu_{0}^{2} +
  \left(\frac{\gamma \epsilon_{1}}{m_{\rm e} c^{2}}\right)^{2} \frac{(1 -
    \beta x_{0})^{2}(1 - \mu_{0})^{2}}{1 - \frac{\gamma
      \epsilon_{1}}{m_{\rm e} c^{2}}(1 - \beta x_{0})(1-\mu_{0})}\right],
\end{multline} 
where 
\begin{equation}
%This is just above Eq.3.33 in Cerutti:2007
x_{0} = \frac{1 - \frac{\epsilon_{0}}{\epsilon_{1}}(1 - \beta \cos
  \theta_{0}) + \frac{\gamma \epsilon_{0}}{m_{\rm e}c^{2}}(1 - \beta
  \cos \theta_{0})(1 + \beta C_{\theta_{0}})}
{\beta + \frac{\gamma \epsilon_{0}}{m_{\rm e}c^{2}}(1 - \beta
  \cos \theta_{0})(\beta + C_{\theta_{0}})},
\end{equation}
and
\begin{equation}
%This is just above Eq.3.33 in Cerutti:2007
K = \frac{\left\{1 - \frac{\gamma \epsilon_{1}}{m_{\rm e}c^{2}}[1 +
    \beta C_{\theta_{0}} - (\beta + C_{\theta_{0}})x_{0}]\right\}^{2}}
{\left| -\beta \gamma \epsilon_{1} - \frac{\epsilon_{1}^{2}}{m_{\rm e}c^{2}}C_{\theta_{0}}\right|}.
\end{equation}
Eq.~\ref{eq:anisotropicICemissivity_singleElectron} is only
valid between the energy limits $\epsilon_{\rm min} \leq \epsilon_{1} \leq
\epsilon_{\rm max}$, where
\begin{equation}
%This is Eq.3.36 in Cerutti:2007
\epsilon_{\rm min/max} = \frac{\gamma m_{\rm e}c^{2}(1 - \beta \cos
  \theta_{0})\epsilon_{0}}
{\gamma m_{\rm e}c^{2} + \epsilon_{0} \pm \sqrt{\epsilon_{0}^{2} +
    m_{\rm e}^{2}c^{4}\gamma^{2}\beta^{2} + 2 \epsilon_{0} \beta
    \gamma m_{\rm e}c^{2}\cos \theta_{0}}}.
\end{equation}

\begin{figure}
\includegraphics[width=8.0cm]{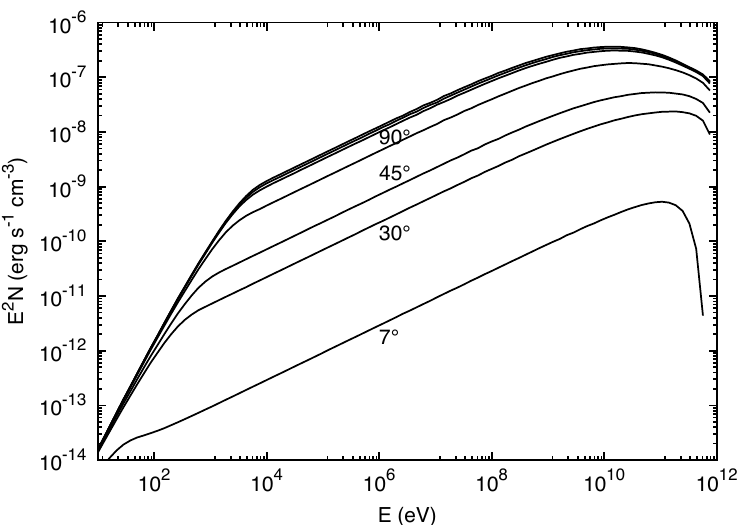}
\caption{The inverse Compton luminosity $L_{\gamma}=E^{2}_{\gamma}
  N_{\gamma}(E_{\gamma}$) calculated for a point-like star and
  different values of the viewing angle $\psi$. The energy
  distribution of the incident photons is a black-body with
  $T=3.9\times10^{4}$\,K. The electron distribution is a power-law
  $N_{\rm e} (E_{\rm e}) \propto E_{\rm e}^{-2}$ over the energy range
  $10\,m_{\rm e}c^{2} \leq E_{\rm e} \leq 1$\,TeV.}
\label{fig:inverseComptonEmissivityWithAngle}
\end{figure}

To obtain the total emission,
Eq.~\ref{eq:anisotropicICemissivity_singleElectron} must be integrated
over the incident photon and electron distributions, the collision
angle, and the volume containing the non-thermal particles, $V$. For
incident photons from a star, the geometry is illustrated in
Fig.~\ref{fig:starElectronObserverGeometry}, which is based on Fig.~4.1
in \citet{Cerutti:2007}. The polar axis $y$ is chosen so that it joins
the centre of the star and the interaction site, which are separated by
a distance $d$. The direction vector for any photon emitted by the
star can be written as
\begin{equation}
%This is just above Eq.4.2 in Cerutti:2007
\vec{e}_{*} = (\sin \alpha \cos \chi, \cos \alpha, \sin \alpha \sin \chi).
\end{equation}
$\chi$ can take the range $0 \leq \chi \leq 2\pi$, but the polar angle
is limited due to the size of the star to the range $0 \leq \alpha
\leq \alpha_{*}$, where $\alpha_{*} = \arcsin (R_{*}/d)$ and $R_{*}$
is the stellar radius. 

If the system is seen with a viewing angle $\psi$, then the scattered
photon has the unit vector
\begin{equation}
%This is just above Eq.4.2 in Cerutti:2007
\vec{e}_{\rm obs} = (\sin \psi, \cos \psi, 0).
\end{equation}
Because the inverse Compton emission from energetic electrons is
highly beamed this is also the direction vector of the electron, $\vec{e}_{e}$. The
collision angle, $\theta_{0}$, can then be obtained from the scalar
product of $\vec{e}_{e}$ and $\vec{e}_{*}$:
\begin{equation}
%This is Eq.4.3 in Cerutti:2007
\vec{e}_{e} \cdot \vec{e}_{*} = \cos \theta_{0} = \cos \psi \cos \alpha
+ \sin \psi \sin \alpha \cos \chi.
\end{equation} 
The resulting emission (${\rm photons/s/erg}$) is given by
\begin{multline}
%This is Eq.4.7 in Cerutti:2007
\frac{dN}{dtd\epsilon_{1}} = \int_{V} dV \int^{E_{\rm e,max}}_{E_{\rm e,min}}  
\int^{\epsilon_{0,{\rm max}}}_{\epsilon_{0,{\rm min}}}
\int^{2 \pi}_{0} \int^{\alpha_{*}}_{0}  
\frac{dN}{dtd\epsilon_{1}}(E_{\rm
  e},\epsilon_{0},\alpha,\chi)\times \\
n_{\rm ph}(\epsilon_{0})N_{\rm e}(E_{\rm e}) \cos \alpha \sin \alpha
d\alpha d\chi d\epsilon_{0} dE_{\rm e},
\end{multline}
where $n_{\rm ph}(\epsilon_{0})$ is the number density of incident
photons at energy $\epsilon_{0}$ (in units of ${\rm
  photons/cm^{3}/erg/sr}$), $N_{\rm
  e}(E_{\rm e})$ is the non-thermal electron distribution (in units of
${\rm electrons/erg/cm^{3}}$), and
$\cos \alpha \sin \alpha d\alpha d\chi$ is the projection of an
element of area $d^{2}S$ on the surface of the star. 
%$N_{\rm e}(p_{\rm e})=4\pi p^{2}f_{\rm pe}(p)$ 

% Anisotropic IC notes: When $\theta_{\rm 0}$ (the angle between the
%direction vector of the relativistic electron and the incoming
%(stellar) photon in the lab frame) equals zero, the scattered photon
%cannot be more energetic than the incoming photon (no energy boost
%takes place). When $\theta_{\rm 0} = \pi$ (i.e. a head-on collision),
%the scattered photon is at least as energetic as the incoming photon
%but can reach significantly higher energies (in the Thomson limit it
%can reach an energy of $4\gamma^{2}\epsilon_{0}$ for $\beta \approx 1$).
%For energetic electrons the inverse Compton emission is highly
%beamed, so most of the radiation that reaches the observer comes from
%electrons moving towards them. [p124]

In Fig.~\ref{fig:inverseComptonEmissivityWithAngle} the variarion of
the inverse Compton emissivity with viewing angle is shown for a
point-like star with a blackbody photon distribution scattering off
non-thermal electrons with a power-law energy distribution. The effect
of the anisotropy is clear to see. Most of the variation is between $0^{\circ}
< \psi < 90^{\circ}$, and there is virtually no change for $135^{\circ} < \psi
< 180^{\circ}$.

\subsection{Relativistic bremsstrahlung emission}
\label{sec:appRB}
The $\gamma$-ray emission (${\rm photons/s/erg}$) from relativistic
bremsstrahlung resulting from the interaction of non-thermal electrons
with thermal protons is \citep[e.g.][]{Blumenthal:1970}
\begin{equation}
q_{\gamma} (E_{\gamma}) = c \int_{V} dV n_{\rm p} \int^{E_{\rm e}^{\rm
    max}}_{E_{\rm e}^{\rm min}} \frac{d\sigma_{\rm Br}}{dE_{\gamma}}(E_{\gamma},E_{\rm
    e}) N_{\rm e}(E_{\rm e}) dE_{e},
\end{equation}
where $E_{\gamma}$ is the photon energy, $n_{\rm p}$ is the number
density of thermal protons, $E_{\rm e}$ is the energy of the
non-thermal electron, and $E_{\rm e}^{\rm max}$ and $E_{\rm e}^{\rm
  min}$ are the maximum and minimum energy of the non-thermal
electrons. The differential cross section (in units of ${\rm
  cm^{2}/erg}$) for the emission of a photon
by a non-thermal electron (with energy $E_{\rm e} >> m_{\rm e}c^{2}$) in the
presence of a proton is \citep[e.g.,][]{Berezinskii:1990}
\begin{multline}
\frac{d\sigma_{\rm Br}}{d E_{\gamma}}(E_{\gamma},E_{e}) = \frac{4
  \alpha_{\rm FS} r_{\rm e}^{2}}{E_{\gamma}} \left[1 + \left(1 -
    \frac{E_{\gamma}}{E_{e}}\right)^{2} - \frac{2}{3}\left(1 -
    \frac{E_{\gamma}}{E_{e}}\right) \right] \times \\
\left\{{\rm ln}\left[\frac{2
        E_{e}(E_{e}-E_{\gamma})}{m_{\rm e}c^{2}E_{\gamma}}\right] -
        \frac{1}{2}\right\},
\end{multline}
where $\alpha_{\rm FS}$ is the fine structure constant.

\subsection{$\pi^{0}$-decay emission}
\label{sec:appPi0}
The $\gamma$-ray emission (${\rm photons/s/erg}$) from the decay of neutral pions is
\begin{equation}
q_{\gamma} (E_{\gamma}) = 2 \int^{E_{\rm p}^{\rm max}}_{E_{\rm
    min}} \frac{Q_{\pi^{0}}(E_{\pi})}{\sqrt{E_{\pi}^{2} -
    m_{\pi^{0}}^{2}c^{4}}} dE_{\pi},
\end{equation} 
where $E_{\pi}$ is the neutral pion energy and
\begin{equation}
E_{\rm min} = E_{\gamma} + \frac{m_{\pi^{0}}^{2}c^{4}}{4 E_{\gamma}}.
\end{equation} 
$Q_{\pi^{0}}$ is the injection function of neutral pions (${\rm
  pions/s/erg}$). In the delta functional approximation it is given by \citep{Aharonian:2000}
\begin{equation}
Q_{\pi^{0}}(E_{\pi}) = \int_{V} dV \frac{\tilde{n}}{K_{\pi}} c n_{\rm p} \sigma_{\rm
  pp}(E_{\rm p}) N_{\rm p}(E_{\rm p}),
\end{equation} 
where $\tilde{n}$ is the number of neutral pions created per
proton-proton collision (it is assumed that $\tilde{n}=1$ and does not
depend on the energy of the proton), $K_{\pi}=0.17$ is the fraction of
the proton kinetic energy that goes into creating the pion, $n_{\rm
  p}$ is the number density of thermal protons, and $N_{\rm p}(E_{\rm p})$ is the non-thermal proton
distribution (in units of ${\rm protons/erg/cm^{3}}$) where the proton
energy is $E_{\rm p}$. The inelastic proton-proton
cross-section, $\sigma_{\rm pp}$, is accurately approximated as
\citep*{Kelner:2006}
\begin{equation}
%See Eq A11 in Vila:2012
\sigma_{\rm pp} (E_{\rm p}) = \left(34.3 + 1.88L +
  0.25L^{2}\right)\left[1 - \left(\frac{E_{\rm
        th}}{E_{p}}\right)^{4}\right]^{2}\,{\rm mb},
\end{equation} 
where $L = {\rm ln}(E_{p}/{\rm 1\,TeV})$ and $E_{\rm th} = (m_{p} + 2m_{\pi} + m_{\pi}^{2}/2 m_{p})c^{2} = 1.22\,$GeV is the threshold
energy for the production of a single $\pi^{0}$.
 
% Don't change these lines
\bsp	% typesetting comment
\label{lastpage}
\end{document}